\documentclass[]{interact}

\usepackage{epstopdf}
\usepackage[doublespacing]{setspace}
\usepackage[longnamesfirst,sort]{natbib}
\bibpunct[, ]{(}{)}{;}{a}{,}{,}

\usepackage{amsfonts,amsmath,amssymb}
\usepackage{amsthm}
\usepackage{graphicx}
\graphicspath{{pictures/}}
\usepackage{color}
\usepackage{pifont}
\usepackage{amsmath}
\usepackage{amsfonts,amsmath,amssymb}
\usepackage{graphicx}
\usepackage{amsbsy}
\graphicspath{{pictures/}}
\usepackage{amsmath}
\usepackage{color,multirow}
\usepackage{algpseudocode}
\usepackage{comment}
\usepackage{mathtools}
\usepackage{complexity}
\usepackage{subcaption}
\usepackage{hyperref}
\usepackage{enumitem}
\usepackage{footnote}
\usepackage{comment}
\usepackage{amsmath}
\usepackage{hyperref}
\usepackage[nameinlink]{cleveref}
\usepackage{natbib}

\usepackage[textsize=scriptsize,textwidth=3cm,shadow]{todonotes}

\newcommand{\plcom}[1]{\todo[color=orange!50!white]{{PL}: #1}}

\theoremstyle{plain}
\newtheorem{theorem}{Theorem}[section]

\theoremstyle{definition}

\newtheorem{example}[theorem]{Example}

\theoremstyle{remark}

\def\arrvline{\hfil\kern\arraycolsep\vline\kern-\arraycolsep\hfilneg}

\newcommand{\new}[1]{{\color{black}#1}}
\newcommand{\neu}[1]{{\color{black}#1}}

\begin{document}


\title{Improving Ranking Quality and Fairness\\ in Swiss-System Chess Tournaments\footnote{A 2-page abstract of this work appeared at The 23rd ACM Conference on Economics and Computation (EC'22).}
}

\author{
\name{Pascal Sauer\textsuperscript{a}, \'{A}gnes Cseh\textsuperscript{b,c}\thanks{CONTACT \'{A}gnes Cseh. Email: agnes.cseh@uni-bayreuth.de}, and Pascal Lenzner\textsuperscript{d}}
\affil{\textsuperscript{a} Potsdam Institute for Climate Impact Research, Potsdam, Germany;\\ \textsuperscript{b} Institute of Economics, HUN-REN Centre for Economic and Regional Studies, Budapest, Hungary;\\ \textsuperscript{c} Department of Mathematics, University of Bayreuth, Bayreuth, Germany;\\ \textsuperscript{d} Hasso Plattner Institute, University of Potsdam, Potsdam, Germany}
}

\maketitle

\begin{abstract}
The International Chess Federation (FIDE) imposes a voluminous and complex set of player pairing criteria in Swiss-system chess tournaments and endorses computer programs that are able to calculate the prescribed pairings. The purpose of these formalities is to ensure that players are paired fairly during the tournament and that the final ranking corresponds to the players' true strength order.

We contest the official FIDE player pairing routine by presenting alternative pairing rules. These can be enforced by computing maximum weight matchings in a carefully designed graph. We demonstrate by extensive experiments that a tournament format using our mechanism (1) yields fairer pairings in the rounds of the tournament and (2) produces a final ranking that reflects the players' true strengths better than the state-of-the-art FIDE pairing system. 
\end{abstract}


\section{Introduction}
How can one determine the relative strength of players who engage in a one-on-one competitive game? This is easy to find out for a group of two players: just let them play a match. For more players, tournaments solve this problem by ranking the players after a limited number of pairwise matches among the participants. The \textit{tournament format} defines a general structure of matches to be played and the method for deriving a ranking from the results of those matches.

\subsubsection*{Tournament Formats}
Most tournaments follow an elimination, a round-robin, or a Swiss-system format. In each round of an \textit{elimination} tournament, such as the second stage of the FIFA World Cup, only players who won their match in the previous round are paired again. The last player standing wins the tournament, and the remaining players' strength can only be estimated very roughly from the round they were eliminated in. \textit{Round-robin} tournaments are also called all-play-all tournaments, because each player plays against each other player once. The player with the highest score at the end of the tournament is declared the winner. The pool stage of the FIFA World Cup consists of round-robin tournaments.

The \emph{Swiss-system} tournament format is widely used in competitive games like most e-sports, badminton, and chess, the last of which this paper focuses on. In such tournaments, the number of rounds is predefined, but the pairing of players in each of these rounds depends on the results of previous rounds. This format offers a convenient golden middle way between the earlier mentioned two tournament formats. However, the features of the Swiss system challenge organizers greatly. Firstly,  unlike in elimination tournaments, the goal is to determine a whole ranking of the players and not only to declare the winner. Secondly, the final ranking of each player is greatly influenced by her assigned opponents, which is not an issue in round-robin tournaments.

Therefore, a mechanism that computes suitable player pairings for Swiss-system tournaments is crucially important. However, designing such a system is a challenging task as it boils down to solving a complex combinatorial optimization problem. Interestingly, the state-of-the-art solution to this problem in chess tournaments relies on a complex set of declarative rules and not on a combinatorial optimization algorithm. In this paper we provide an algorithmic approach and we demonstrate that it outperforms the declarative state-of-the-art solution. For this, we do not try to mimic the FIDE solution but instead focus on the most important features of the Swiss system and derive a maximum weight matching formulation that enforces them.

\subsubsection*{The Swiss-System in Chess}
In Swiss-system chess tournaments, there are two well-defined and rigid \textit{absolute} and two milder \textit{quality} pairing criteria. \neu{These criteria form the backbone of the much more specific and rigid declarative FIDE rules~\cite[Chapter C.04]{fide2020handbook}.}
\begin{enumerate}
\item[(A1)] No two players play against each other more than once.
\item[(A2)] In each round before the last one, the difference of matches played with white 
and matches played with black pieces is between $-2$ and $2$ for every player.
\item[(Q1)] Opponents have equal or similar score. 
\item[(Q2)] Each player has a balanced color distribution.
\end{enumerate}
Criterion (A1) ensures variety, while criterion (A2) ensures fairness, since the player with white pieces starts the game, and thus has an advantage over her opponent~\citep{Hen92,milvang2016prob}. These absolute criteria must be obeyed at any cost, which often enforces the relaxation of the two quality criteria.

In order to implement criterion~(Q1), players with equal score are grouped into \textit{score groups}. In each round, a chosen \textit{pairing system} allocates each player an opponent from the same score group. If a complete pairing is not possible within a score group, then one or more players are moved to another score group. Criterion~(Q2) requires that after each round of the tournament, the difference between matches played with black and white pieces is small for each player.

Adhering to these four criteria makes pairing design truly challenging. \neu{Besides incorporating the four criteria, the FIDE also requires that the same pairing is generated whenever the same situation arises at competitions. To ensure this uniqueness, additional rigid declarative rules have been added~\cite[Chapter C.04]{fide2020handbook}.} Pairings at FIDE tournaments were traditionally calculated manually by so-called arbiters, often using trial-and-error. Today, pairings are computed by decision-making software, but the FIDE pairing criteria are still written for human instead of computer execution. Over the years, more and more criteria were added to resolve ambiguities, which increased the complexity to a level at which pairing decisions are very challenging to comprehend for most players and even arbiters.

\subsection{Related Literature} 
Novel algorithms that assist tournament scheduling regularly evoke interest in the AI and Economics communities~\citep{HI03,LJCS14,BEM15,KW15,CIT16,DS18,GRSZ18,Hos18,Kar18,VG19,LGS22}. Also, due to its relation to voting, analyzing tournament solutions, mostly variants of round-robin tournaments, is a prominent research direction in Economics and Social Choice Theory~\citep{moulin1986choosing,laslier1997tournament,BrandtF07,hudry2009survey,StantonW11,brandt2016introduction,kim2017can,SaileS18,brandt2018structure}.
We first elaborate on existing work on tournament formats, and then turn to approaches that utilize matchings for scheduling tournaments. \new{For the reader not familiar with graphs and combinatorial optimization, we suggest to consult the book of \cite{KV12}.}

\subsubsection*{Comparing Tournament Formats}
\citet{appleton1995may} gives an overview of tournament formats and compares them with respect to how often the best player wins. 
\citet{scarf2009numerical} simulate different tournament formats using team data from the UEFA Champions League.
Recently, a comparative study by \citet{sziklai2021efficacy} found that the Swiss-system tournament is the most efficient format in terms of approximating the true ranking of the players.
Moreover, \citet{elmenreich2009robustness} compare several sorting algorithms, including one based on a Swiss-system tournament, with respect to their robustness, which is defined as the degree of similarity between the resulting ranking and the true strength order of players. They find round-robin sort, merge sort, and Swiss-system sort to be the most robust overall.

\subsubsection*{Swiss-System Tournaments}
Sports tournaments are by far not the only application area of the Swiss system. Self-organizing systems~\citep{FW09}, person identification using AI methods~\citep{WTW+15}, and choosing the best-fitting head-related transfer functions for a natural auditory perception in virtual reality~\citep{OVF19} are all areas where the Swiss system appears as a solution concept. To the best of our knowledge, there are only a few papers that analyze Swiss-system tournaments. The works of \citet{csato2013ranking,Csato17,Csato21} study the ranking quality of real-world Swiss-system tournaments, in particular, whether based on the match results a fairer ranking could have been obtained by different scoring rules. However, this research is orthogonal to our approach since pairing rules are not considered.


\subsubsection*{Automated Matching Approaches} A tournament schedule can be seen as a set of matchings---one for each round. 
\citet{glickman2005adaptive} propose an algorithm based on maximum weight perfect matchings to find the schedule. This algorithm maximizes the information gain about players' skill. The authors' approach compares favorably against random and Swiss-system pairing if at least 16 rounds are played. However, almost all real-world Swiss-system chess tournaments have less than 10 rounds according to~\url{chess-results.com} \citep{herzog2020chess}.

\citet{kujansuu1999stable} use the stable roommates problem, see \citet{irving1985efficient}, to model a Swiss-system tournament pairing decision. Each player $p$ has a preference list, which ranks the other players by how desirable a match between player $p$ and each other player would be. The desirability depends on score difference and color balance. In comparison to the official FIDE pairing, this approach produces pairings with slightly better color balance but higher score differences between paired players, or, in other words, clearly favors criterion~(Q2) over~(Q1).

\subsubsection*{Weighted Matching Models for Chess Tournaments} The two papers closest to ours focus on modeling the exact FIDE pairing criteria and computing the prescribed pairings. 

\citet{olafsson1990weighted} pairs players using a maximum weight matching algorithm on a graph, where players and possible matches are represented by vertices and edges. Edge weights are set so that they model the 1985 FIDE pairing criteria. At that time, pairing criteria were more ambiguous than today, and pairing was done by hand, which sometimes took several hours. In contrast, using {\'O}lafsson's method, pairings could be calculated fast. Pairings calculated with the commercial software built by {\'O}lafsson are claimed to be preferred by experts to manually calculated pairings. However, {\'O}lafsson only provides examples and does not present any comparison based on formal criteria.

A more recent attempt to convert the FIDE pairing criteria into a weighted matching instance was undertaken by 
\citet{BFP17}. Due to the extensive criterion system, only a subset of the criteria were modeled. The authors show that a pairing respecting these selected criteria can be calculated in polynomial time, and leave it as a challenging open question whether the other FIDE criteria can also be integrated into a single weighted matching model. The contribution appears to be purely theoretical, since neither a comparison with other pairing programs, nor implementation details are provided.

Our work breaks the line of research that attempts to implement the declarative FIDE pairing criteria via weighted matchings. Instead, we design new pairing rules along with a 
different mechanism to compute the pairings, and demonstrate their superiority compared to the FIDE pairing criteria and engine. This clearly differentiates our approach from the one in~\citet{olafsson1990weighted,BFP17}.

\subsection{Preliminaries and FIDE Criteria}\label{chap:background}

\textit{Players} are entities participating in a Swiss-system tournament. Each player has an \emph{Elo rating}, which is a measure designed to capture her current playing strength from the outcome of her earlier matches \citep{elo1978rating}. In a \textit{match} two players, $a$ and $b$, play against each other. The three possible \textit{match results} are: $a$ wins and $b$ loses, $a$ and $b$ draw, $a$ loses and $b$ wins. The winner receives 1 point, the loser 0 points, while a draw is worth 0.5 points. A Swiss-system tournament consists of multiple \textit{rounds}, each of which is defined by a \textit{pairing}: a set of disjoint pairs of players, where each pair plays a match. At the end of the tournament, a strict ranking of the players is derived from the match results.

\subsubsection*{Bye Allocation} In general, each player plays exactly one match per round. For an odd number of players, one of them receives a so-called `bye', which is a point rewarded without a match. This is always the player currently ranked last among those who have not yet received a bye.

\subsubsection*{Color Balance}
The FIDE Handbook~\cite[Chapter C.04.1]{fide2020handbook} states that `For each player the difference between the number of black and the number of white games shall not be greater than 2 or less than -2.' This criterion may only be relaxed in the last round. This corresponds to our criterion (A2). 
Also, a ban on a color that is assigned to a player three times consecutively, and further milder criteria are phrased to ensure a color assignment as close to an alternating white-black sequence as possible \cite[Chapters C.04.3.A.6 and C.04.3.C]{fide2020handbook}. \new{The color assignment in the first round is drawn randomly.}

\subsubsection*{Pairing Systems}
\label{sec:pairing_systems}
Players are always ranked by their current tournament score. Furthermore, within each score group the players are ranked by their Elo rank. The score groups and this ranking are the input of the \emph{pairing system}, which assigns an opponent to each player. Three main pairing systems are defined for chess tournaments. Table~\ref{tab:pairing_systems_example_pairing} shows an example pairing for each of them.
\begin{itemize}
\item \textbf{Dutch:}
Each score group is cut into an upper and a lower half. The upper half is then paired against the lower half so that the $i$th ranked player in the upper half plays against the $i$th ranked player in the lower half. Dutch is the de facto standard for major chess tournaments.
\item \textbf{Burstein:}
For each score group, the highest ranked unpaired player is paired against the lowest ranked unpaired player repeatedly until all players are paired.
\item \textbf{Monrad:}
In ascending rank order each unpaired player in a score group is paired against the next highest ranked player in that score group.
\end{itemize}

\new{The vast majority of chess tournaments use the Dutch system, however,  Burstein is the main pairing principle behind the team pairings at the prestigious Chess Olympiads~\citep{fide2023_olympiad}, while Monrad is commonly used in Denmark and Norway~\citep{Wikipedia2023monrad}}.

\begin{table}[ht]
\centering
\begin{tabular}{ccccccc}
Dutch   &&& Burstein    &&& Monrad\\
1--5     &&& 1--8         &&& 1--2\\
2--6     &&& 2--7         &&& 3--4\\
3--7     &&& 3--6         &&& 5--6\\
4--8     &&& 4--5         &&& 7--8\\
\end{tabular}
\caption{Example pairing for each pairing system in a score group of 8 players. Players are referenced by rank within the score group, i.e., player 1 has the highest Elo rank.}
\label{tab:pairing_systems_example_pairing}
\end{table}
\noindent For comparison, we propose two additional pairing systems based on randomness.
\begin{itemize}
    \item \textbf{Random:} Every player within a score group is paired against a random player from her score group.
    \item \textbf{Random2:} Every player from the top half of her score group is paired against a random player from the bottom half of her score group.
\end{itemize}

\subsubsection*{Floating Players}
Players who are paired outside of their own score group are called~\emph{floaters}. To ensure that opponents are of similar strength--our criterion~(Q1)--, the FIDE criteria require to minimize the number of such floaters and aim to float them to a score group of similar score. However, 
floating is unavoidable, e.g., in score groups with an odd number of players, and also in score groups where the first or second criterion eliminates too many possible matches.

\subsubsection*{The BBP Pairing Engine}
\label{sec:bbp_engine}
A \textit{pairing engine} is used to calculate the pairing for each round, based on the results of previous rounds. The BBP pairing engine was developed by 
\citet{bierema2017bbp}. It implements the FIDE criteria strictly \cite[C.04.3 and C.04.4.2]{fide2020handbook} for the Dutch and Burstein pairing systems and outputs the unique pairing adhering to each of them. BBP uses a weighted matching algorithm, similarly as the approaches in \citet{olafsson1990weighted,BFP17}. The main difference to our algorithm is that while the weighted model of BBP was designed to follow the declarative criteria of FIDE and output the prescribed pairings, our pairing engine relies on a different weighted model, computes completely different pairs, and while doing so, it is able to reach a better ranking quality and a higher degree of fairness. The output of Dutch BBP will serve as a base for our comparisons throughout the paper, because Dutch is the sole pairing system implemented by  programs currently endorsed by the FIDE \cite[C.04.A.10.Annex-3]{fide2020handbook} and because the BBP pairing engine is open-source. We further remark that even though a Burstein BBP code exists and has been made public, its author draws attention to the fact that it is ``a flawed implementation of a version of the Burstein system, not endorsed by the FIDE Systems of Pairings and Programs Commission.''

\subsubsection*{Final Ranking}
The major organizing principle for the final ranking of players is obviously the final score. Players with the same final score are sorted by tiebreakers. The FIDE \cite[Chapter C.02.13]{fide2020handbook} defines 14 types of tiebreakers, and the tournament organizer lists some of them to be used at the specific tournament. If all tiebreaks fail, the tie is required to be broken by drawing of lots.

\subsection{Our Contribution}
In this paper, we present a novel mechanism for calculating pairings in Swiss-system chess tournaments. With this, we contest the state-of-the-art \new{FIDE pairing criteria, which are implemented by the BBP pairing engine. } 
We compare the two systems by three measures: ranking quality, number of floaters, and color balance quality, in accordance with the FIDE tournament schedule goals. 
Our main findings are summarized in the following list \new{and in Table~\ref{ta:cont}}.
\begin{enumerate}
    \setlength\itemsep{0mm}
    \item  We implemented the pairing systems Dutch, Burstein, Monrad, Random, and Random2 with an extensible and easy-to-understand approach that uses maximum weight matchings.
    \item The pairing systems in descending order by expected \textbf{ranking quality} are: Burstein $>$ Random2 $>$ Dutch $=$ Dutch BBP $>$ Random $>$ Monrad. In particular, our implementations of Burstein and Random2 both yield higher ranking quality, while our implementation of Dutch yields similar ranking quality as the one reached by the Dutch BBP pairing engine.
   \item We utilize our weighted matching model to define a novel measure called `normalized strength difference', which we identify as the main reason for a good ranking quality. This also explains why our approach outperforms the Dutch BBP engine.   
   \item The pairing systems in ascending order by expected \textbf{number of floaters} are: Burstein $<$ Random2 $=$ Dutch $=$ Monrad $<$ Dutch BBP $<$ Random. Compared to Dutch BBP, \new{almost all variants of} our mechanism \new{are} fairer in terms of matching more players within their own score group.
    \item All our pairing systems ensure the same \textbf{color balance quality} as Dutch BBP, with Random even reaching a better color balance. Moreover, we show that our approach can easily be modified to enforce an even stronger color balance. This does not significantly affect the ranking quality---only the number of floaters increases slightly.  
\end{enumerate}
\begin{quote}
\new{
\textbf{Main Take-Away:} Our new implementations of Dutch, Burstein, and Random2 either outperform or are on a par with Dutch BBP in all measured aspects. Thus, we propose to use our implementations as a pairing engine in future FIDE chess tournaments.}
\end{quote}

\begin{table}[ht]
\begin{center}
\resizebox{\textwidth}{!}{
\begin{tabular}{|l|llllll|}
	      \noalign{\hrule}
       Ranking quality & Burstein \arrvline & Random2 \arrvline & Dutch & \textcolor{blue}{Dutch BBP} \arrvline & Random \arrvline & Monrad\\
       \noalign{\hrule}
       Number of floaters & Burstein \arrvline& Random2 & Dutch & Monrad \arrvline & \textcolor{blue}{Dutch BBP} \arrvline & Random\\
       \noalign{\hrule}
       Color balance & Random \arrvline & Random2 & Dutch & \textcolor{blue}{Dutch BBP} & Burstein & Monrad\\
	      \noalign{\hrule}
\end{tabular}
    }
\end{center}
\caption{The hierarchy of the discussed six pairing systems, with the benchmark Dutch BBP being marked blue. Each row represents a metric, and the earlier a system appears in the row, the better it performs measured with the corresponding metric. Blocks in rows denote ties.}
\label{ta:cont}
\end{table}


\noindent Rankings derived from pairwise comparisons constitute an active research area, 
mainly due to their versatile applicability, extending from web search and recommender systems~\citep{CBC+13,BCD+19} to sports tournaments~\citep{Sin88,csato2013ranking}. For simplicity and reproducibility, we focus on the latter application area, but our approach has the potential to be applied to other areas as well, where pairwise comparison schedules are designed.

\section{Pairings via Maximum Weight Matching}
\label{sec:mwm_engine}

Our novel mechanism is based on computing a maximum weight matching (MWM) in an auxiliary, suitably weighted graph. The MWM engine is optimized for simplicity: score groups, color balances, and the employed pairing system are modeled by weights, so only a single computation of a MWM is needed in each round. We now describe the MWM engine.

\subsection{Input}

Each tournament has $n$ players $P = \{p_1,\dots,p_n\}$, a chosen pairing system (Dutch, Burstein, Monrad, Random, or Random2), and a maximum allowed color difference~$\beta$. As criterion~(A2) states, FIDE aims for $\beta=2$. If $n$ is odd, the weakest performing player who has not received a bye yet is given one, in accordance with the FIDE rules. In the MWM engine we will exclude the same player while constructing the auxiliary graph. Hence, from this point on we can assume that $n$ is even.

\noindent Before each tournament round, the following input parameters are defined for each player $p_i \in P$:
\begin{itemize}
    \item $Elo(p_i)$: the Elo rating of $p_i$ prior to the tournament. This remains unchanged for all rounds.
    \item $s(p_i)$: the current score of $p_i$, defined as the sum of points player $p_i$ collected so far.
    \item $r(p_i)$: the current rank of $p_i$, calculated from ordering all players in decreasing order according to their scores and their Elo ratings. Higher score and higher Elo rating yield better rank. Players with equal Elo rating are ordered randomly at the beginning, and their order is kept for all rounds.
    \item $cd(p_i)$: the current color difference of $p_i$, defined as the number of matches played with white minus the number of matches played with black pieces.
\end{itemize} 

\subsection{Graph Construction}
With these parameters as input, we construct the corresponding auxiliary weighted graph $G_r = (V,E,w)$ for round $r$ as follows. Let $V := P$ and for all pairs of players $p_i\neq p_j$, let the edge set $E$ contain the edge $\{p_i,p_j\}$ if 
\begin{enumerate}
 \item[(1)] $p_i$ and $p_j$ have not yet played against each other, and
 \item[(2)] $|cd(p_i) + cd(p_j)| < 2\beta$.
\end{enumerate}
These rules ensure criteria~(A1) and (A2). The second condition in our model will enforce $-2 \leq cd(p_i) \leq 2$ together with our color assignment rule in Section~\ref{sec:alg}. In Section~\ref{sec:betterbalance} we additionally consider a variant where $-1 \leq cd(p_i) \leq 1$ is enforced. This implements FIDE's criterion that the color assignment should be as close to an alternating white-black sequence as possible and that no player can be assigned the same color three times in a row.

The weight of an edge $\{p_i,p_j\} \in E$ is defined as the tuple $$w(p_i,p_j) := (-|s(p_i)-s(p_j)|, -|cd(p_i) + cd(p_j)|, \pi(p_i,p_j)),$$ where the value of $\pi(p_i,p_j)$ depends on the pairing system as follows.
\begin{itemize}
    \item Monrad: $\pi(p_i,p_j) :=-\left| r(p_i)-r(p_j) \right|$.
    \item Burstein: $\pi(p_i,p_j) :=\left| r(p_i)-r(p_j) \right|^{1.01}$.
    \item Dutch:  $\pi(p_i,p_j) :=-\left|\frac{\text{sg size}}{2}- |r(p_i)-r(p_j)| \right|^{1.01}$, where sg size  is set to 0 if $p_i$ and $p_j$ belong to different score groups, and it is the size of the score group of $p_i$ and $p_j$ otherwise.
    \item Random: $\pi(p_i,p_j) :=$ random number in the interval $(0,1)$.
    \item Random2: $\pi(p_i,p_j)$ is set to a random number in the interval $(0,1)$ if $p_i$ and $p_j$ belong to different halves of the same score group, otherwise it is set to a random number in the interval $(-1,0)$.
\end{itemize}
The exponent 1.01 in the function for Burstein rewards a larger rank difference, i.e., the Burstein pairing in Table~\ref{tab:pairing_systems_example_pairing} indeed carries a larger weight than the Dutch pairing, which has the same sum of rank differences. Similarly, the exponent for Dutch penalizes a larger distance from $\frac{\text{sg size}}{2}$. Notice that this exponent could be an arbitrary number as long as it is larger than~1.

\begin{figure}[ht]
 \centering
 \includegraphics[width=0.85\textwidth]{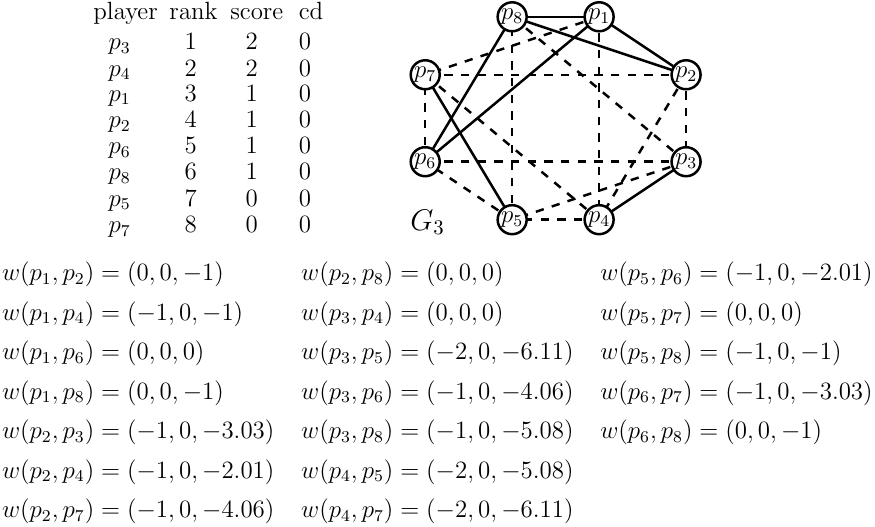}
 \caption{A sample instance of the graph construction for a tournament with eight players. 
 The current ranking at the tournament is shown on the top left, the constructed graph is shown top right and its edge weights that arise from using the Dutch pairing system are depicted on the bottom.
 Edge weight values are rounded. Bold edges are possible matches within the same score group whereas dashed edges are other possible matches. Missing edges are matches that were already played or that are forbidden due to the color balance criterion. For example, the edge weight $w(p_1,p_8) = (-|s(p_1)-s(p_8)|, -|cd(p_1) + cd(p_8)|, \pi(p_1,p_8)) = (-|1-1|,-|0+0|,-\left|\text{sg size}/2 - |r(p_1)-r(p_8)| \right|^{1.01}) = (0,0,-\left|4/2 - |3-6| \right|^{1.01}) = (0,0,-\left|2 - 1 \right|^{1.01}) = (0,0,-1)$.
 The instance corresponds to round 3 of the example presented in 
 Example~\ref{algo_example}. There, also the corresponding maximum weight matching consisting of all edges with weight $(0,0,0)$ is shown.}
 \label{fig:example_weights}
\end{figure}

See Figure~\ref{fig:example_weights} for an illustration that shows the corresponding edge weights using the Dutch pairing system for a sample instance.

\subsection{Algorithm}
\label{sec:alg}
The edge weights of $G_r$ are compared lexicographically and a maximum weight matching is sought for. This implies that pairing players within their score groups has the highest priority, optimizing color balance is second, and adhering to the pairing system is last. The comprehensive rules of our framework consist of our two absolute rules for including an edge in  
the graph $G_r$, and this priority ordering serving as our quality rule. 

Before round $r$, we compute a maximum weight matching $M$ in graph $G_r$ and derive the player pairing from the edges in~$M$. If $\{p_i,p_j\} \in M$ then the players $p_i$ and $p_j$ will play against each other in round~$r$. Between them, the respective player with the lower color difference will play white. If they have the same color difference---e.g., in the first round---, then colors are assigned randomly.

\begin{example}\label{algo_example}
 We consider pairings of an example 4-round tournament with 8 players generated via the MWM engine using the Dutch pairing system. 
 
 Initially players are sorted decreasingly according to their Elo rating. In the following figures, bold edges are possible matches within the same score group, whereas dashed edges are other possible matches. The maximum weight matchings are shown in red. Arrows within the tables indicate the match outcomes (winner points to loser, no draws), and the color column shows the corresponding color distribution. The table for round $i+1$ is based on the table of round $i$. 

 \begin{figure}[ht]
 \centering
 \includegraphics[width=0.6\textwidth]{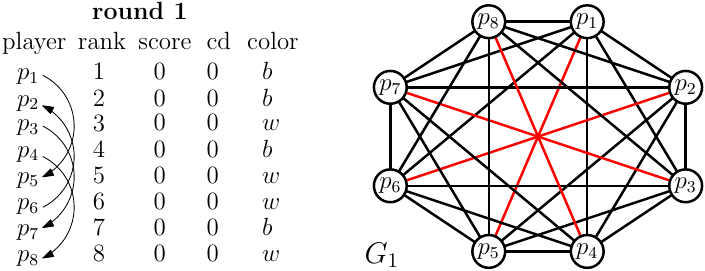}
 \caption{Round 1 pairing of the example tournament. }
 \label{fig:example1}
 \end{figure}

  \begin{figure}[ht]
 \centering
 \includegraphics[width=0.6\textwidth]{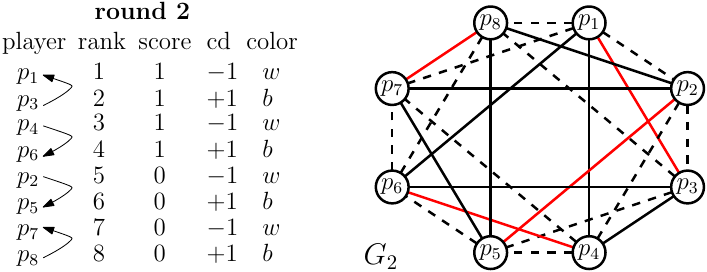}
 \caption{Round 2 pairing of the example tournament. }
 \label{fig:example2}
 \end{figure}
 
 \begin{itemize}
     \item As score and color difference are equal, the pairing in round 1 is enforced by the Dutch pairing system. See Figure~\ref{fig:example1}.
     
 \item The pairing in round~2, depicted in Figure~\ref{fig:example2}, is the outcome of optimizing first for criterion~(Q1) and then for criterion~(Q2), e.g., in $G_2$ we have $w(p_1,p_3) = w(p_4,p_6) = (0,0,-1)$ and $w(p_1,p_4) = w(p_3,p_6) = (0,-2,0)$ so the MWM chooses the edges $\{p_1,p_3\}$ and $\{p_4,p_6\}$.
 
 \item In round 3, depicted in Figure~\ref{fig:example3}, in $G_3$ players $p_3$ and $p_4$ are paired since $w(p_3,p_4) = (0,0,0)$ whereas the weight of any other incident edge of both $p_3$ and $p_4$ has lexicographically lower weight.

 \item Finally, the round 4 matching in $G_4$, depicted in Figure~\ref{fig:example4}, is enforced by maximizing the number of matches within score groups. If $p_1$ and $p_2$ would be paired, then, since $p_3$ and $p_4$ already played, player $p_4$ would float to a match with a player with score $1$, which implies that no match within the group with score $1$ is possible.
 \end{itemize}

 \begin{figure}[ht]
 \centering
 \includegraphics[width=0.6\textwidth]{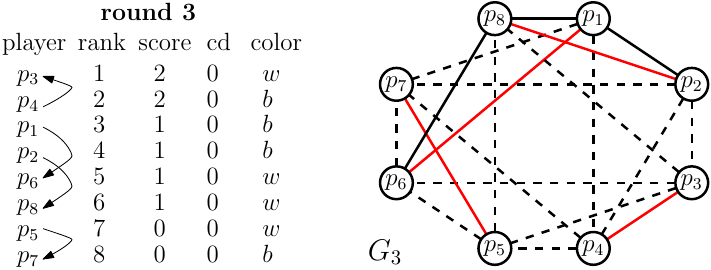}
 \caption{Round 3 pairing of the example tournament. }
 \label{fig:example3}
 \end{figure}
 
 \begin{figure}[ht]
 \centering
 \includegraphics[width=0.6\textwidth]{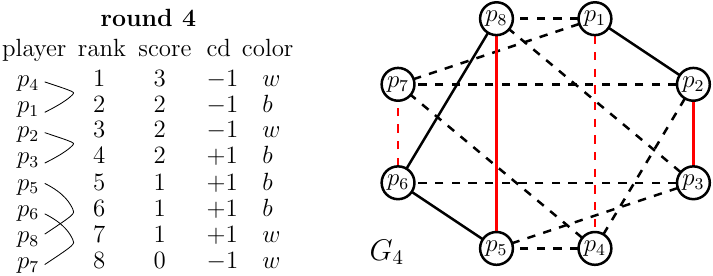}
 \caption{Round 4 pairing of the example tournament. }
 \label{fig:example4}
 \end{figure}
\end{example}

\begin{example}\label{bbp_example}
 For comparison, we now apply Dutch BBP to the same instance from Example~\ref{algo_example} and display the calculated player pairings in Figure~\ref{fig:BBPexample}. Match results in the final round are not displayed, as they do not influence the pairings. Even though we copied as many match results as possible from Example~\ref{algo_example}, the two engines calculate largely different player pairings.
 
 \begin{figure}[h]
 \centering
 \includegraphics[width=\textwidth]{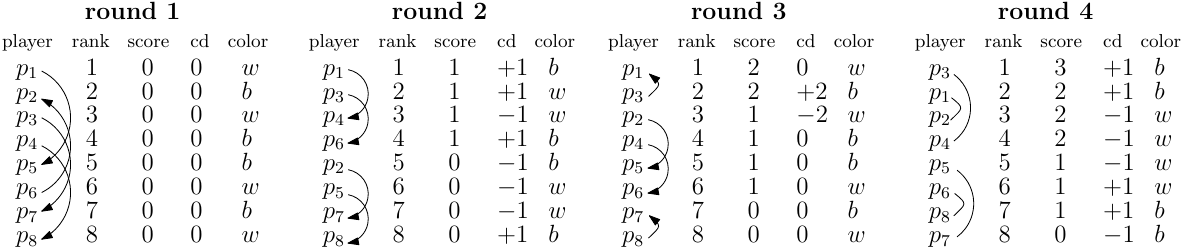}
 \caption{Player pairings for the tournament from Example~\ref{algo_example} calculated via Dutch BBP. Match results are consistent with the corresponding results from Example~\ref{algo_example}.}
 \label{fig:BBPexample}
 \end{figure}
 The pairing in the first round is solely determined by the main Dutch pairing principle, and thus is identical to the pairing produced by our Dutch MWM engine. However, the color assignment is different, as in the match between $p_1$ and $p_5$, Dutch BBP assigns white to $p_1$, whereas Dutch MWM assigns white to~$p_5$. This is to be expected, as both engines are supposed to draw the first-round color assignment randomly---however, as our tests with Dutch BBP revealed, the initial color assignment there is deterministic. 
 In the first round, we gave the exact same match results to Dutch BBP as to Dutch MWM. Despite of this, from the second round on, player pairings calculated by the two engines are completely different. For those pairs that do not appear in Example~\ref{algo_example}, we calculated the outcome from Milvang's probability distribution \citep{milvang2016prob}, otherwise we kept the same match result as in Example~\ref{algo_example}. 

 A notable difference of both approaches for the same sample tournament is that Dutch MWM overall achieves a better color balance than Dutch BBP. In the latter one, we have color differences of $2$ in round 3 whereas the color differences in Dutch MWM never exceed 1. In Dutch MWM the players perfectly alternate between playing with white and black pieces.
 

\end{example}

\section{Assumptions and Experimental Setup}
In our simulations we assume that each player $p_i \in P$ has true playing strength $str(p_i)$ that is approximated by her Elo rating $Elo(p_i)$ and we treat both values as constant throughout the tournament. \new{It is crucial that $str(p_i)$ and $Elo(p_i)$ might differ, as they are used in different components of our model, which we describe in the next two paragraphs. 
} 

The probabilities of match results and optimal rankings are defined by the playing strength. More precisely, each player's playing strength is a random number drawn from an uniform distribution of values between 1400 and 2200. We also justified our claims on ranking quality using other realistic player strength distributions. We elaborate on these in the appendix. The results are in line with the results for the uniform distribution.

Elo ratings are used for computing $r(p_i)$ and for breaking ties in the final order. The Elo rating of player $p_i$ is randomly drawn from a normal distribution with mean $str(p_i)$ and standard deviation $\frac{3000-str(p_i)}{20}$. This function mirrors the assumption that a higher Elo rating estimates the strength more accurately. 

To avoid the noise introduced by byes, we assume that the number of players $n$ is even. The number of rounds is chosen to lie between $\lceil\log_2 n \rceil$ and $\frac{n}{2}$, as at least $\lceil\log_2 n \rceil$ rounds ensure that a player who wins all matches is the sole winner and at most $\frac{n}{2}$ rounds ensures that, according to Dirac's theorem \citep{dirac1952some}, a perfect matching always exists. The tiebreakers used for obtaining the final tournament ranking are based on the FIDE recommendation \cite[C.02.13.16.5]{fide2020handbook}.

\subsubsection*{Computing the Maximum Weight Matching}
First we transform each edge weight given as a tuple to a rational number. In particular, $w(p_i,p_j)$ is transformed to $10\,000 \cdot (-|s(p_i)-s(p_j)|) + 100 \cdot (-|cd(p_i) + cd(p_j)|)+ \pi(p_i,p_j)$. The factors 10\,000 and 100 ensure that each lexicographically maximum solution corresponds to a maximum weight solution with the new weights and vice versa. We compute pairings using the LEMON Graph Library \citep{dezsHo2011lemon} implementation of the maximum weight perfect matching algorithm, which is based on the blossom algorithm of 
\citet{edmonds1965paths} and has the same time and space complexity \citep{Kol09}. The implementation we use has $O(nm \log{n})$ time complexity, where $n$ is the number of players and $m$ is the number of edges in the constructed graph~$G_r$. 

\subsubsection*{Realistic Probabilistic Model for Match Results}\label{sec:probabilistic_model}
The results of the individual matches are computed via a probabilistic model that is designed to be as realistic as possible. Match results are drawn at random from a suitably chosen probability distribution based on the players' strength and on the assigned colors for that match. For this, we use the probability distribution proposed by 
\citet{milvang2016prob}, which was featured in a recent news article of the FIDE commission System of Pairings and Programs \citep{fide2020news}. Milvang's probability distribution was engineered via a Data Science approach that used real-world data from almost 4 million real chess matches from 50\,000 tournaments. It is based on Elo ratings and color information, whereas we use true strength values instead of Elo ratings to get unbiased match result probabilities. 

Using Milvang's approach, the probability for a certain outcome of a match depends on the actual strengths of the involved players, not only on their strength difference. Draw probability increases with mean strength of the players. The probabilities also depend on colors, as the player playing with white pieces has an advantage. See Table~\ref{tab:example_probabilities} for some example values drawn from Milvang's distribution. 
\begin{table}[ht]
\centering
\begin{tabular}{rccc}
Player Strengths   & Win White  & Win Black & Draw\\
1200 (w) vs 1400 (b) & 26 \%& 57 \%& 17 \%\\
2200 (w) vs 2400 (b) & 14 \%& 55 \%&31 \%\\
2400 (w) vs 2200 (b) & 63 \%& 11 \%& 26 \% 
\end{tabular}
\caption{Example match outcome probabilities drawn from Milvang's probability distribution \citep{milvang2016prob}.
}
\label{tab:example_probabilities}
\end{table}

\subsubsection*{Measuring Ranking Quality} Ranking quality measures how similar the tournament's final ranking is to the ranking that sorts the players by their strength. One popular measure for the difference between two rankings is the Kendall $\tau$ distance \citep{kendall1945treatment}. It counts the number of discordant pairs: pairs of elements $x$ and $y$, where $x < y$ in one ranking, but $y < x$ in the other. We use its normalized variant, where $\tau \in [-1, 1]$, and $\tau = 1$ means the rankings are identical, while $\tau = -1$ means one ranking is the inverse of the other. A higher Kendall $\tau$ is better, because it indicates a larger degree of similarity between the true and the output ranking.

We also justify our claims on ranking quality using two other well-known and possibly more sophisticated similarity measures, the Spearman $\rho$ distance \citep{spearman1904proof} and normalized discounted cumulative gain (NDCG). We elaborate on these measures and their behavior for our problem in the appendix. The results are in line with the ones derived for the Kendall $\tau$ distance.

\subsubsection*{Measuring Fairness}
We measure fairness in terms of the two relaxable criteria of Swiss-system chess tournaments: (Q1) on the equal score of opponents and (Q2) on the color distribution balance. Adhering to (Q1) is measured by the number of float pairs, which equals the number of matches with opponents from different score groups throughout the tournament. We measure the absolute color difference of a round as the sum of color differences for all players:  $acd = \sum_{p_i \in P}{|cd(p_i)|}$. Note that as $|cd(p_i)| \geq 1$ for all players after each odd round, $acd \geq n$ in those rounds.

\subsubsection*{Presentation of the Data}
Data is presented in the form of \textit{violin plots} \citep{hintze1998violin}, \textit{letter value plots} \citep{HWK17}, and \emph{scatter plots} \citep{FD05}. For violin plots, kernel density estimation is used to show a smoothed probability density function of the underlying distribution. Additionally, similar to box plots, quartiles are shown by dashed lines. Letter value plots are enhanced box plots that show more quantiles. Unlike violin plots, they are suitable for discrete values, as all shown values are actual observations without smoothing.

Our plots compare the MWM implementation of the five pairing systems with the BBP implementation of Dutch.

\section{Simulation Results}
All simulations use the following parameters, unless noted otherwise:
\begin{itemize}
    \item number of players $n$: 32
    \item number of rounds: 7
    \item strength range: between 1400 and 2200
    \item maximum allowed color difference $\beta$: 2 
    \item sample size: 100\,000
\end{itemize}
These values were chosen to be as realistic as possible, based on parameters of more than 320\,000 real-world tournaments uploaded to the website \url{chess-results.com}.\footnote{The data was kindly provided by Heinz Herzog, author of the FIDE-endorsed tournament manager \url{Swiss-Manager} \citep{herzog2020swiss} and \url{chess-results.com} \citep{herzog2020chess}.} The experiments were run on a computer server using version 20.04.1 of the Ubuntu operating system. It is powered by 48 Intel Xeon Gold 5118 CPUs running at 2.3 GHz and 62.4 GiB of RAM. We emphasize that the standard real-life challenge at a tournament, that is, 
computing a single pairing via a maximum weight matching for a tournament round can be solved 
in a fraction of a second on a standard laptop.

\subsection{Ranking Quality}
\label{sec:renking_q}
The pairing system of a Swiss-system tournament has a major impact on the obtained ranking quality, as Figure~\ref{fig:ranking_quality} shows. Burstein and Random2 achieve the best ranking quality, followed by Dutch and Dutch BBP. Random has a worse ranking quality and Monrad performs by far the worst. For other strength ranges, Figure~\ref{fig:mean_strength} shows consistent results. \new{See also Tables~\ref{table:mean_values} and~\ref{table:median_values} for the corresponding mean and median Kendall $\tau$ values.}
\begin{figure}[ht]
    \centering
    \includegraphics[width=0.7\linewidth]{color_figures/ranking_quality_pairing_systems
    .pdf}
    \caption{Ranking quality measured by normalized Kendall~$\tau$. A higher value means a better ranking quality.}
    \label{fig:ranking_quality}
\end{figure} 

\begin{figure}[ht]
    \centering
    \includegraphics[width=0.8\linewidth]{color_figures/mean_strength
    .pdf}
    \caption{Ranking quality measured by normalized Kendall~$\tau$ for different strength ranges.}
    \label{fig:mean_strength}
\end{figure}

\begin{table}[h]\centering
\begin{tabular}{c|c|c|c|c}
Strength Range & Burstein & Dutch BBP & Dutch    & Random2\\
\hline
1000 -- 1800   & 0.624 & 0.586  & 0.588 & 0.607\\
\hline
1400 -- 2200   & 0.671 & 0.633  & 0.634 & 0.654\\
\hline
1800 -- 2600   & 0.721 & 0.685  & 0.686 & 0.706 \\
\end{tabular}
\caption{\new{Mean normalized Kendall $\tau$ values averaged over 100\,000 simulated tournaments for each configuration as in Figure~\ref{fig:mean_strength}.}}
\label{table:mean_values}
\end{table}
\begin{table}[h]\centering
\begin{tabular}{c|c|c|c|c}
Strength Range & Burstein & Dutch BBP & Dutch    & Random2\\
\hline
1000 -- 1800   & 0.629 & 0.590 & 0.591 & 0.610\\
\hline
1400 -- 2200   & 0.673 & 0.634 & 0.637 & 0.657 \\
\hline
1800 -- 2600   & 0.723 & 0.688 & 0.690 & 0.710 \\
\end{tabular}
\caption{\new{Median normalized Kendall $\tau$ values averaged over 100\,000 simulated tournaments for each configuration as in Figure~\ref{fig:mean_strength}.}}
\label{table:median_values}
\end{table}

\noindent Comparing Dutch to Dutch BBP shows that they behave very similarly, with slight advantage for Dutch. This is remarkable, since Dutch BBP is based on complex and rigid declarative criteria that are time-tested, while Dutch is the output of our easy-to-understand, purely matching-based approach. 
Together with the performance of Burstein and Random2 this shows that more transparent pairing systems can outperform the state-of-the-art Dutch BBP in terms of ranking quality.

We provide additional experimental results on the ranking quality in the appendix. There we present consistent results also for fewer or more players, for other strength range sizes, and for different player strength distributions. 

\subsection{Reasons for High Ranking Quality}\label{sec:reasons}

Here we elaborate on how our flexible maximum weight matching model enables us to detect the exact reason why certain pairing systems produce better rankings, which might help designing better pairing systems in the future. In particular, we provide experiments that shed light on why Burstein, Random2, Dutch, and Dutch BBP reach a better ranking quality than Random and Monrad and why Burstein and Random2 outperform Dutch BBP. 

\begin{figure}[ht]
    \centering
    \begin{subfigure}{0.45\textwidth}
    \includegraphics[width=\textwidth]{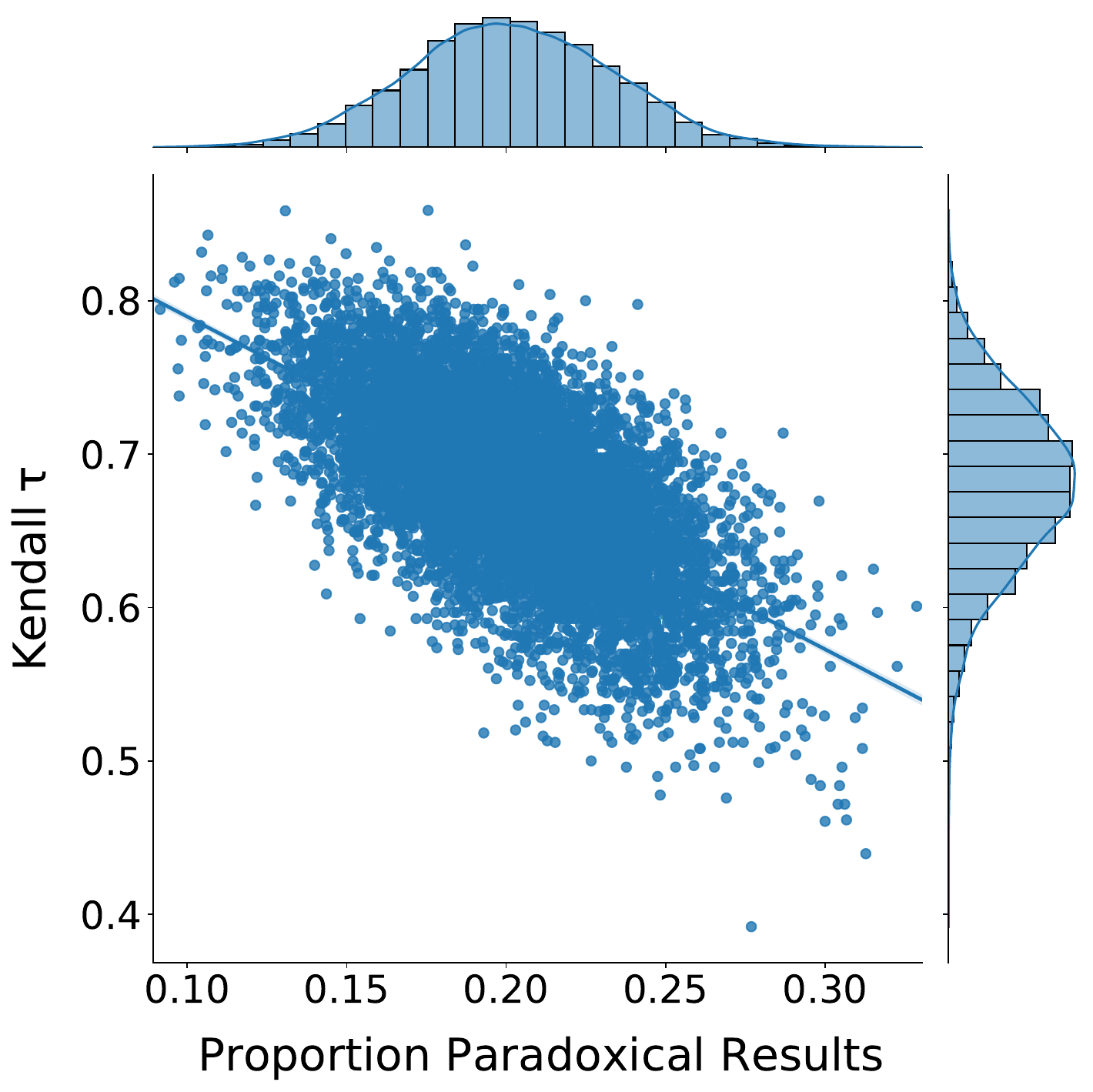}
    \caption{Correlation between unforeseen results and normalized Kendall $\tau$.}
    \label{fig:ppr_rq_after_7}
    \end{subfigure}
    \hfill
    \begin{subfigure}{0.45\textwidth}
    \includegraphics[width=\textwidth]{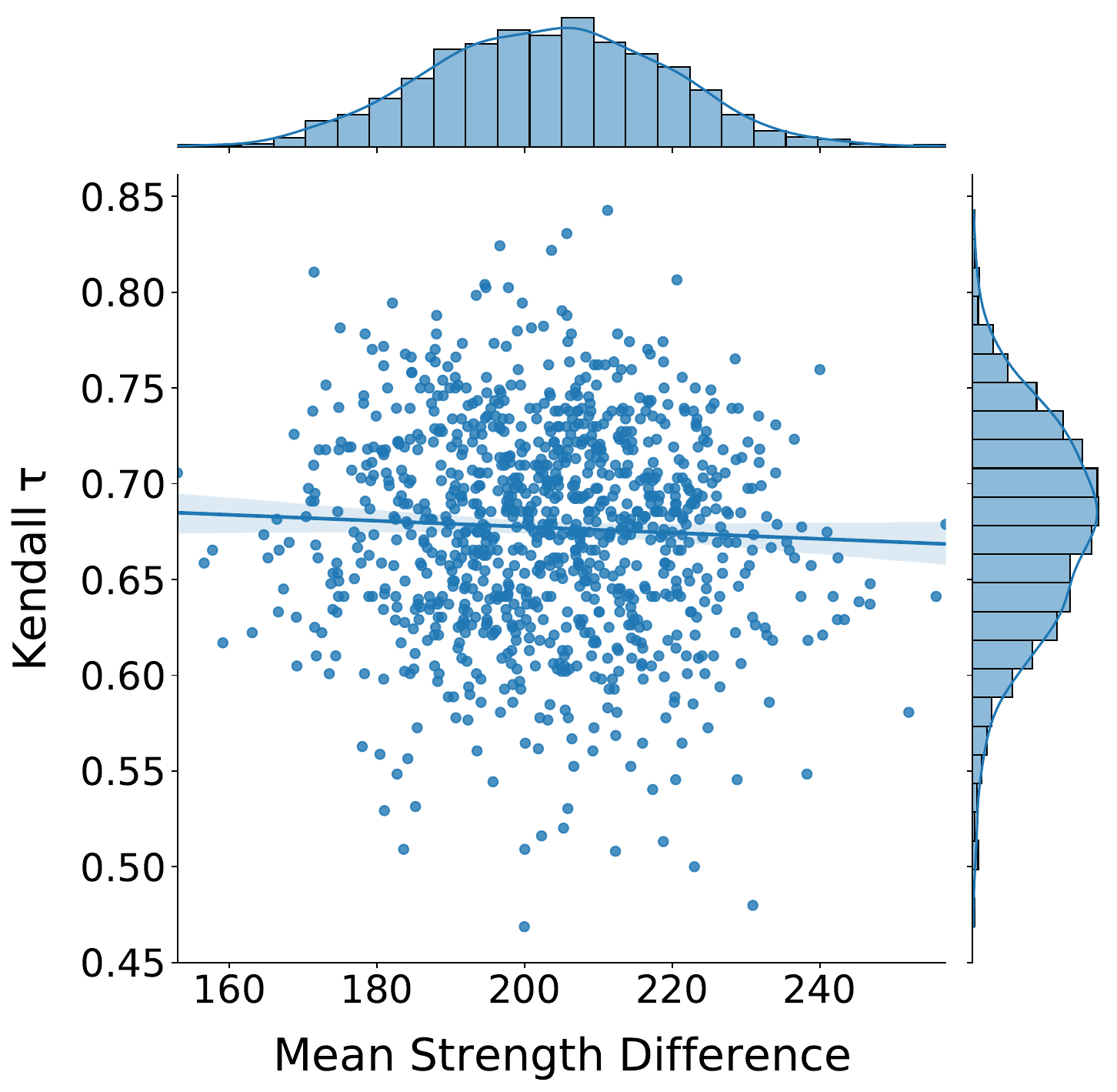}
    \caption{Correlation between mean strength difference and normalized Kendall $\tau$.}
    \label{fig:msd_rq_after_7}
    \end{subfigure}
    \caption{Observed correlations after seven rounds, paired with Dutch BBP.}
    \label{fig:test}
\end{figure}

In order to rank players correctly, their relative playing strength must be approximated from match results. 
We call a match result \textit{unforeseen} if a weaker player wins against a stronger opponent. Unforeseen match results hinder the approximation of both players' strengths, so pairing systems should aim to minimize the number of unforeseen match results. Figure~\ref{fig:ppr_rq_after_7} confirms this,
by showing a strong negative correlation between the proportion of unforeseen results and ranking quality for Dutch BBP. A similar correlation can be observed for all pairing systems.

The probability of an unforeseen match result increases as the strength difference of paired players decreases because the outcome of those matches is less predictable. In general, a higher mean strength difference in a tournament lowers the number of unforeseen match results, which then leads to better ranking quality. Our results in Section~\ref{sec:renking_q} justify the observation that mean strength difference seems to be positively correlated with ranking quality, as mean strength difference is low when using Monrad, medium with Random, and high for Burstein, Random2 and Dutch/Dutch~BBP.

However, when looking at results from Dutch BBP only, there is a small negative correlation instead, as Figure~\ref{fig:msd_rq_after_7} shows.
This is also true for Dutch, Burstein, and Random2. This seemingly unforeseen correlation can be explained as follows. A better ranking leads to a smaller mean strength difference for these pairing systems. In an optimal ranking, each player is in her correct score group, together with players of similar strength, so the mean strength difference will be low. However, in a suboptimal ranking, some players are in a score group that does not reflect their strength. Therefore, these players are either stronger or weaker compared to the other players in their score group, which results in higher mean strength difference.

\begin{figure}[ht]
    \centering
    \includegraphics[width=0.7\linewidth, trim={0 0 0 2cm}, clip]{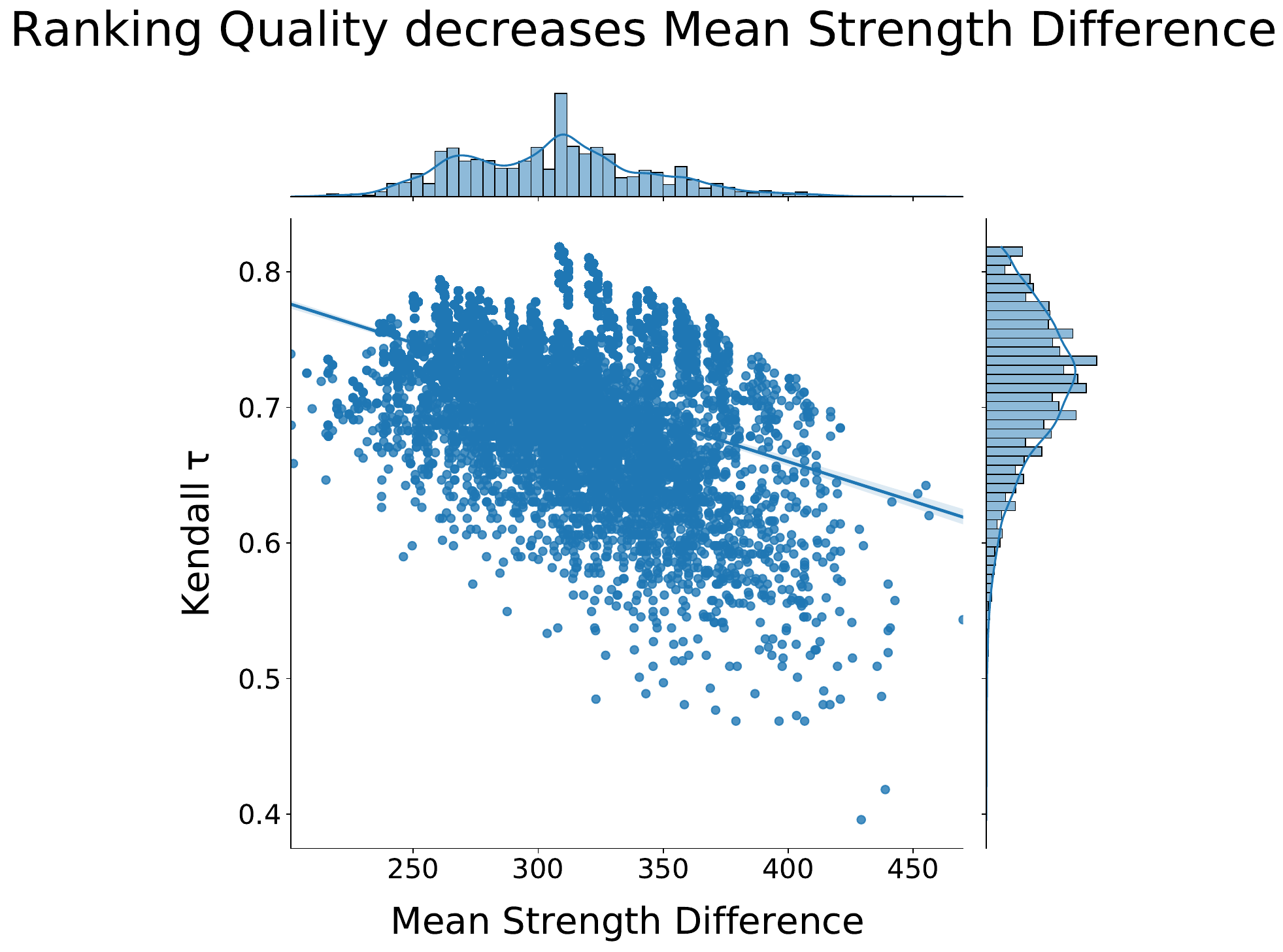}
    \includegraphics[width=0.9\linewidth]{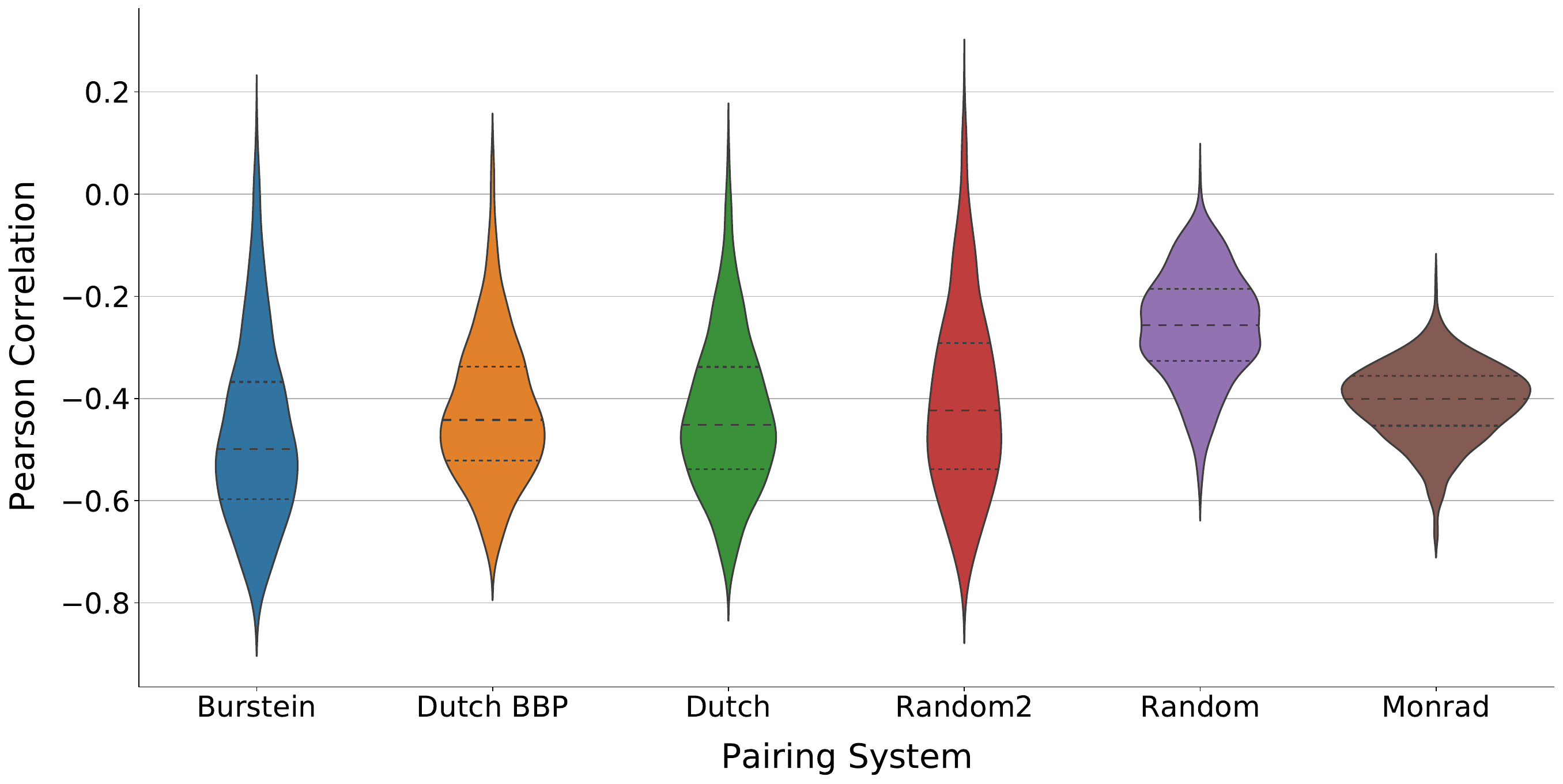}
    \caption{The scatter plot (top) shows the correlation between normalized Kendall $\tau$ and mean strength difference. The violin plot (bottom) shows the distribution of Pearson correlation coefficients if that experiment is repeated for 1\,000 different tournaments, whose first round was simulated 1\,000 times.
    }
    \label{fig:higher_rq_leads_to_lower_msd}
\end{figure}

Figure~\ref{fig:higher_rq_leads_to_lower_msd} shows empirical evidence for this effect: the pairing in round one is always the same, but unforeseen match results due to randomness lead to different rankings, which then determine the mean strength difference in round two.
In our experiment, the same single randomly paired first round was played 10\,000 times. Each time the ranking quality after round one and the mean strength difference of the Dutch BBP pairing for round two was recorded. For the analysis, we use the \textit{Pearson correlation coefficient} that is a standard measure for the linear dependence between two variables. In our case, a negative Pearson correlation coefficient indicates that on average, a higher Kendall $\tau$ is observed together with a lower mean strength difference.

The problem with mean strength difference is that it does not take into account whether a low mean strength difference was the result of a pairing system's choice or due to unfavorable score groups. This can be avoided by taking the maximum possible strength difference into account. 
For this, we define the \textit{normalized strength difference} as the total strength difference divided by the maximum possible total strength difference. 

For computing the normalized strength difference it is essential to calculate the maximum possible strength difference. For this, we again use our maximum weight matching approach, but this time with a pairing system that maximizes strength difference. In particular, we use a modification of our Burstein edge weights $w(p_i,p_j)$ where we set $\pi(p_i,p_j) := |str(p_i)-str(p_j)|$. 
Remember that $str(p_i)$ and $str(p_j)$ are the true strength values of players $i$ and $j$, respectively. Of course, this new pairing system requires knowledge of all true player strengths, so it cannot be used in realistic settings. We only use it as an analytical tool.

\begin{figure}[ht]
    \centering
    \includegraphics[width=0.5\linewidth]{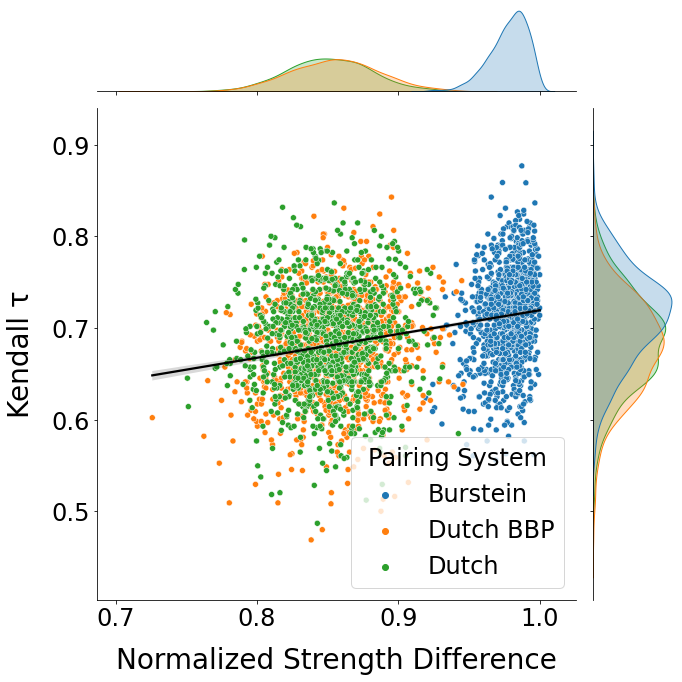}
    \caption{Correlation between ranking quality and normalized strength difference for Burstein, Dutch, and Dutch BBP after seven rounds.}
    \label{fig:bd_nsd}
\end{figure}

Figure~\ref{fig:bd_nsd} compares the normalized strength difference for Dutch BBP and for our maximum weight matching implementation of Burstein and Dutch, where our version of Burstein clearly beats Dutch BBP in ranking quality.
Firstly, this figure shows a positive correlation between normalized strength difference and normalized Kendall $\tau$ for each of Burstein, Dutch, and Dutch BBP after seven rounds. Simulations with each of Random2, Random, and Monrad also indicate a similar positive correlation. Secondly, Figure~\ref{fig:bd_nsd} also demonstrates the positive correlation between the normalized strength difference and the ranking quality across pairing systems. In particular, Burstein clearly beats Dutch BBP in normalized strength difference and also in ranking quality. This correlation is true in general: considering all pairing systems, exactly the ones with a high normalized strength difference (Burstein, Random2, Dutch, Dutch BBP) lead to a good ranking quality, while the ones with medium and low normalized strength difference (Random and Monrad) lead to medium and low ranking quality. 

To summarize, our flexible maximum weight matching model enabled us to detect the exact reason why certain pairing systems produce better rankings. Our surprising finding is that even though at first sight, a high mean strength difference seems to be the pivotal factor, it is actually a high normalized strength difference that results in a better ranking quality. This discovery might help designing better pairing systems in the future.

\subsection{Fairness}
The highly complex pairing criteria of the FIDE were designed with a focus on two fairness goals phrased as quality criteria, (Q1): minimizing the number of float pairs and (Q2): minimizing the absolute color difference. 

\begin{figure}[!ht]
    \centering
    \includegraphics[width=0.7\linewidth]{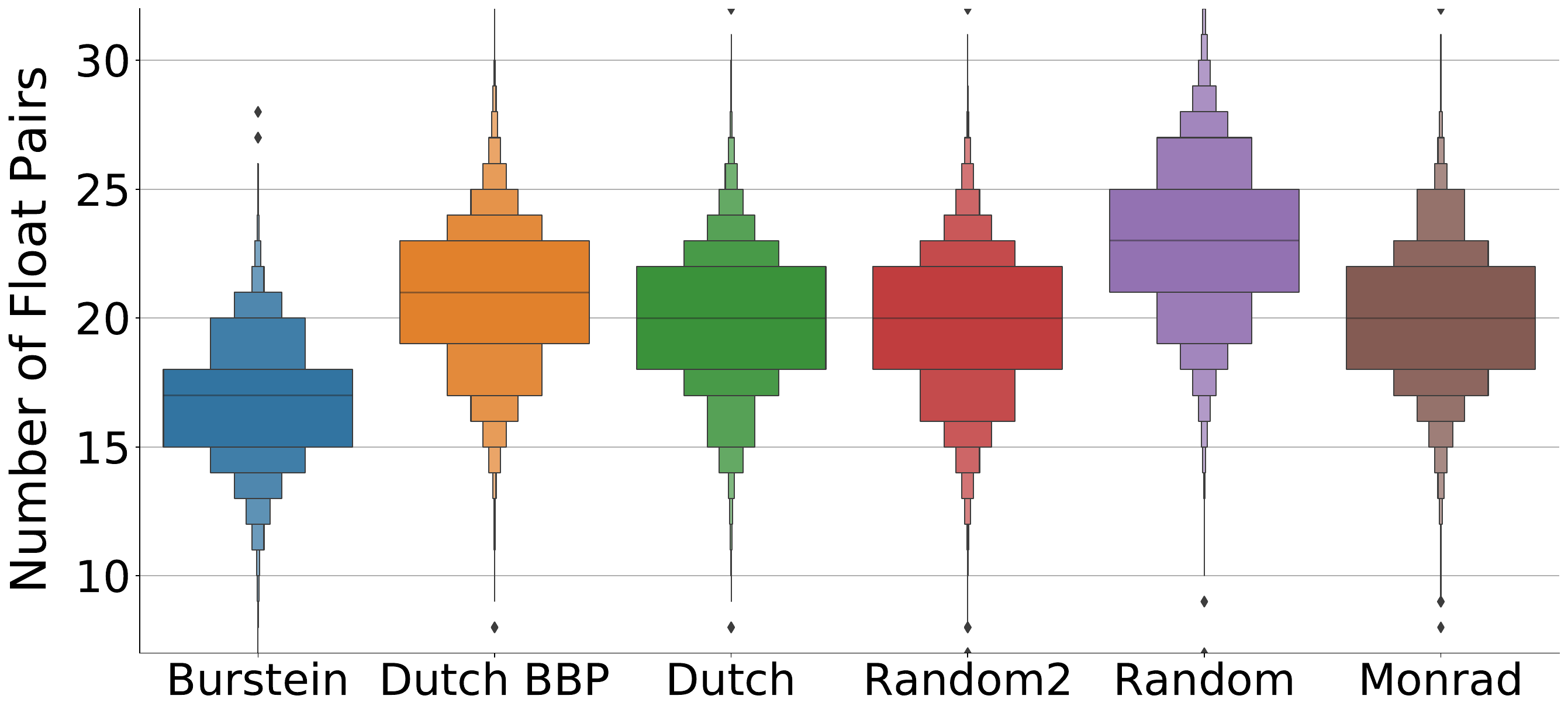}
    \caption{Number of float pairs out of the $7\cdot 16 = 112$ matches of the tournament. Recall that floating is often unavoidable due to the size of the score group. A lower number indicates a  better implementation of criterion~(Q1).}
    \label{fig:float_pairs}
\end{figure}

Criterion~(Q1) is at the heart of Swiss-system tournaments as pairing players of equal score ensures well-balanced matches. As Figure~\ref{fig:float_pairs} shows, Burstein, Dutch, and Random2 beat Dutch BBP in terms of the number of float pairs. In the appendix we show consistent results for other simulation parameters. 

\begin{figure}[!ht]
    \centering
    \includegraphics[width=0.7\linewidth]{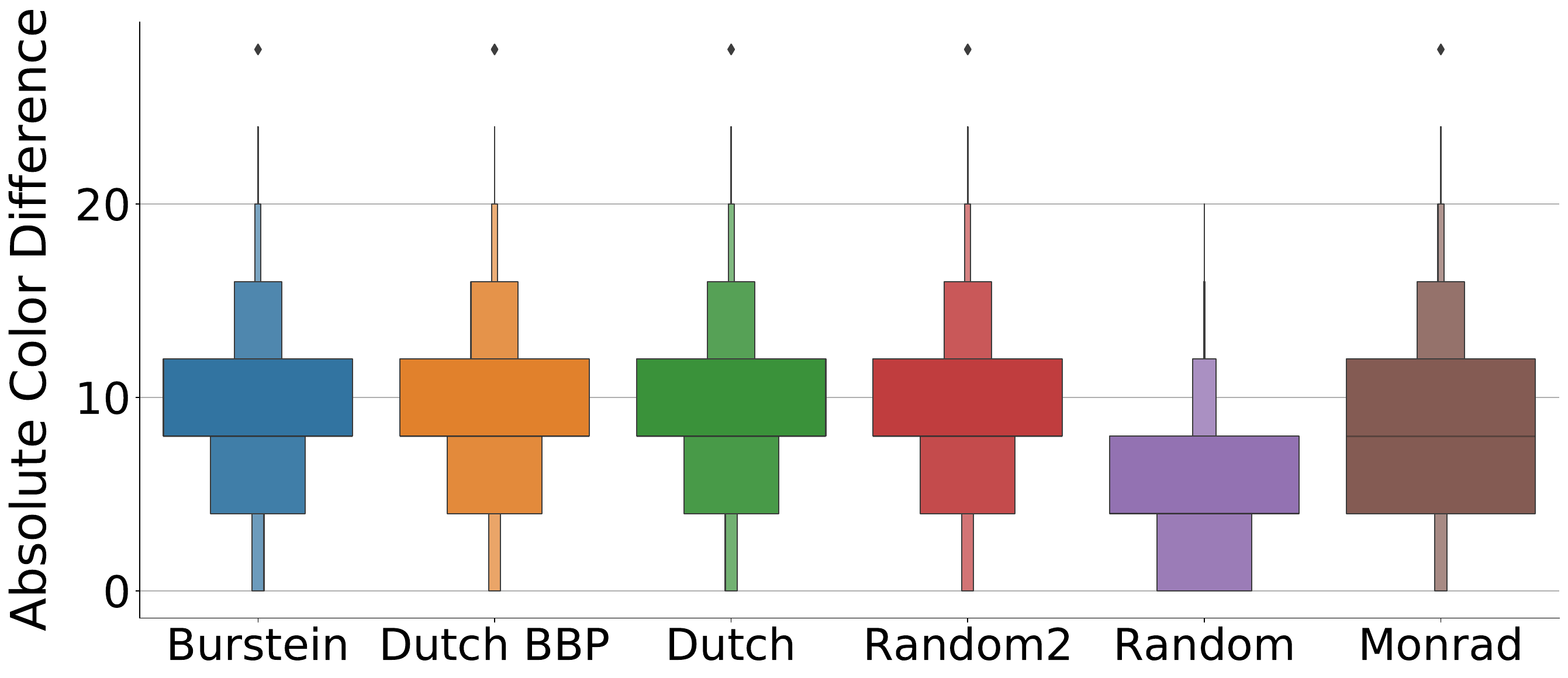}
    \caption{Absolute color difference after 6 rounds. A lower $acd$ means a better color distribution. Recall that a $acd \geq n$ for each odd round, while $acd = 0$ is possible after each even round.}
    \label{fig:color_difference_6}
\end{figure}

\noindent Figure~\ref{fig:color_difference_6} focuses on criterion~(Q2) and shows that an absolute color difference very similar to the one guaranteed by Dutch BBP can be achieved via our MWM engine. The pairing system Random even outperforms Dutch BBP in this regard. In the appendix, we provide additional experiments with different numbers of rounds and numbers of players that lead to consistent results. 

\noindent Hence, our maximum weight matching approach with edge weights that prioritize matches within score groups and secondly optimize for color balance is on a par with the sophisticated official FIDE criteria for criterion~(Q2) and it even outperforms them for criterion~(Q1). Thus, our more transparent approach ensures the same color balance quality but achieves even fewer float pairs. Moreover, our approach also allows for a different trade-off between criteria (Q1) and (Q2) that does not affect the obtained ranking quality.   

\subsection{
Lower Maximum Allowed Color Difference}\label{sec:betterbalance}
So far, for all our experiments we assumed that the maximum allowed color difference $\beta$ equals $2$, i.e., the difference of the number of matches played with white pieces and the number of matches played with black pieces is at most~$2$. This is in line with the official FIDE rules. However, due to the flexibility of our maximum weight matching approach, we can easily enforce an even stronger color difference constraint and observe the impact on the obtained ranking quality and the number of float pairs.

\begin{figure}[ht]
    \centering
    \includegraphics[trim={0 0 0 1600mm},clip,width=0.7\linewidth]{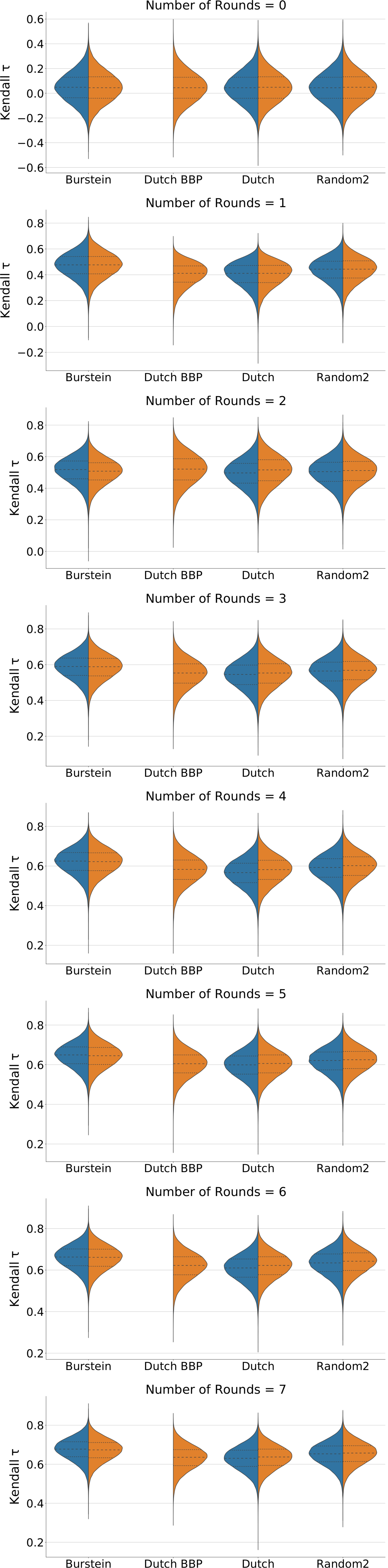}
    \caption{Ranking quality measured by normalized Kendall $\tau$. Results for $\beta=0.1$ are shown in blue, results for $\beta=2$ in orange.}
    \label{fig:color_difference_limit}
\end{figure}

Interestingly, as Figure~\ref{fig:color_difference_limit} shows, the obtained ranking quality is almost the same even if we look at the extreme case of setting $\beta=0.1$, which is equivalent to enforcing an alternating black-white sequence for all players. Notice that setting $\beta$ to anything in the interval $(0,0.5]$ implies that the absolute color difference is $0$ for all even rounds and $n$ for all odd rounds.

\begin{figure}[ht]
    \centering
    \includegraphics[trim={0 0 0 1155mm},clip,width=0.7\linewidth]{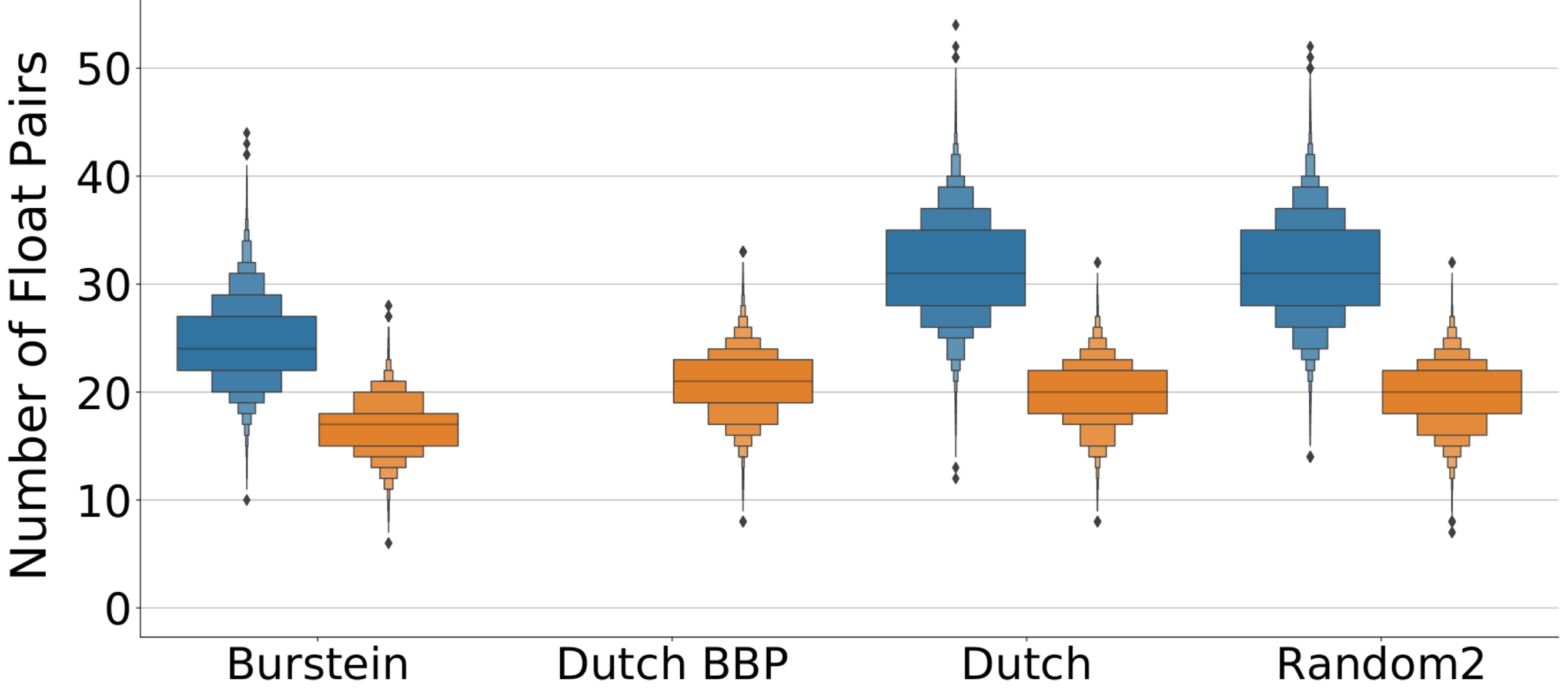}
    \caption{Number of float pairs for 7 rounds. Results for $\beta=0.1$ are shown in blue, results for $\beta=2$ in orange.}
    \label{fig:num_float_pairs_limit}
\end{figure}

Naturally, the high ranking quality for a much more restricted $\beta$ comes at a cost, which can be measured in the number of float pairs. The obtained number of float pairs is influenced by the maximum allowed color difference $\beta$, because for higher $\beta$ it is easier to fulfill criterion (Q1), i.e., to find suitable matches within the corresponding score group. In our experiments we investigate the increase in the number of float pairs when we assume the extreme case of $\beta=0.1$. Figure~\ref{fig:num_float_pairs_limit} shows that the number of float pairs increases for all pairing systems, compared to the case with $\beta=2$. However, the increase is only moderate. This result offers a novel trade-off for tournament organizers: when using the MWM engine, they have the choice between keeping the number of floaters down at the cost of a standard color difference, as advised by FIDE, or they opt for slightly more float pairs, but can guarantee an alternating white-black color assignment for each player. The ranking quality is equally high in both variants.

\section{Conclusion}
The experimental results of our MWM engine with Burstein or Random2 demonstrate that it is possible to outperform the state-of-the-art FIDE pairing criteria in terms of both ranking quality and fairness, i.e., criteria (Q1) and (Q2), with a novel efficient mechanism that is more transparent and intelligible to all participants. The direct comparison of our MWM Dutch engine versus Dutch BBP shows that even if the same pairing system is used,  
MWM achieves the same ranking quality but is more powerful since it yields an improved fairness. We believe that the key to this is the direct formulation of the most important criteria as a maximum weight matching problem.

The only scenario for which we might advise against using our mechanism is when the arbiter has no access to a computing device. In order to manually produce pairings in our framework, the arbiter would need to calculate the edge weights and then execute Edmonds' blossom algorithm. Instead, the FIDE~\cite[Chapter C.04.3.D]{fide2020handbook} provides manually executable rules.
However, these rules include exhaustive search routines that can make the execution very slow, i.e., highly exponential in the number of players \citep{BFP17}. Therefore, the ill-fated arbiter has to choose between learning to execute Edmonds' blossom algorithm and following a cumbersome exponential-time pairing routine. We remark that this latter routine is complex to the point that even pairing engines already endorsed by the FIDE make mistakes occasionally~\citep{BBP22}.

A clear advantage of our mechanism is that it is easily extendable: as Random and Random2 already demonstrate, a new pairing system can be implemented simply by specifying how edge weights are calculated. Similarly, as we have also demonstrated, the color balance can be adjusted by simply changing the parameter~$\beta$. By thinning out the edge set in our graph, we can also reach an alternating black-white sequence for each player instead of just minimizing the color difference in each round. 

The flexibility of the maximum weight matching approach proved to be essential for uncovering the driving force behind the achieved high ranking quality: the normalized strength difference. Hence, our approach was not only valuable for computing better pairings but also in the analysis of the obtained ranking quality. 
Furthermore, the flexibility of the MWM engine likely allows to incorporate additional quality criteria like measuring fairness via the average opponent ratings. Also quality criteria of other games and sports tournaments organized in the Swiss system can be integrated into the model.

Last but not least, other fields using rankings derived from pairwise comparisons might also benefit from our work. A possible application area for our MWM approach is the computation of pairings in a speed-dating type event. Also there a sequence of matchings must be computed, no pair should be matched repeatedly, and the goal is to match like-minded participants. Such events are organized at conferences and other networking occasions~\citep{PN20}, and also resemble mentor assignment preparation at universities~\citep{MP11,GSK+16} and trainee rotation schedules in medicine~\citep{CV20}.


\bibliographystyle{ACM-Reference-Format}
\bibliography{references}

\newpage

\appendix

\section{Ranking Quality}
In the following we discuss additional simulation experiments that measure the obtained ranking quality for various parameter settings.
\subsection{Different Tournament Sizes}
We start with 
experimental results 
demonstrating that our findings on the ranking quality 
remain valid for tournaments 
of different sizes in terms of number of players and number of rounds. 

Usually it is expected that a player who wins all matches also wins the tournament, without being tied for the first place. This can be ensured by playing at least $\lceil \log_2 n \rceil$ rounds: four rounds for 16 players, five rounds for 32 players and six rounds for 64 players. Most tournaments are five or seven rounds long, according to data from chess-results.com \citep{herzog2020chess}.

\begin{figure}[ht]
    \centering
    \includegraphics[trim={0 0 115mm 220mm},clip,width=0.7\linewidth]{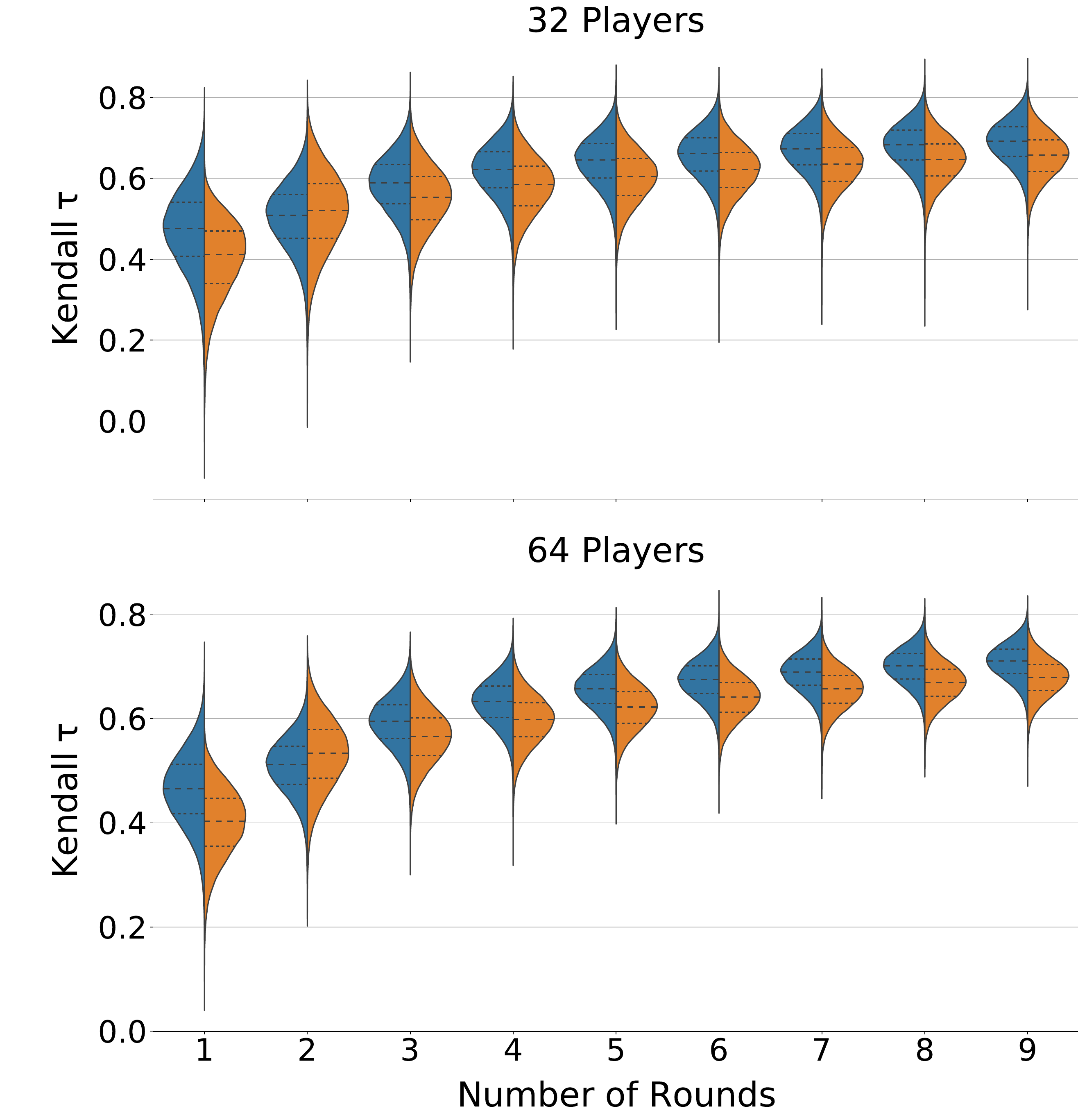}
    \caption{Ranking quality after 1-9 rounds, 32 or 64 players with strength range 1400-2200. Results for Burstein are shown in blue, Dutch BBP results are shown in orange.}
    \label{fig:ranking_quality_number_of_players_rounds}
\end{figure}

In general, more rounds lead to higher ranking quality, although with diminishing effect, as Figure~\ref{fig:ranking_quality_number_of_players_rounds} shows. In terms of the achieved ranking quality,
the MWM engine with Burstein outperforms Dutch BBP in all cases, except for the unrealistic case of a tournament with only two rounds.

\subsection{Different Strength Range Sizes}
Here we vary the used strength range size, i.e., we sample the player strengths from different intervals. A smaller strength range size corresponds to a tournament among players with similar strength and larger strength range sizes model tournaments with more heterogeneous players. 
\begin{figure}[ht]
    \centering
    \includegraphics[width=0.8\linewidth]{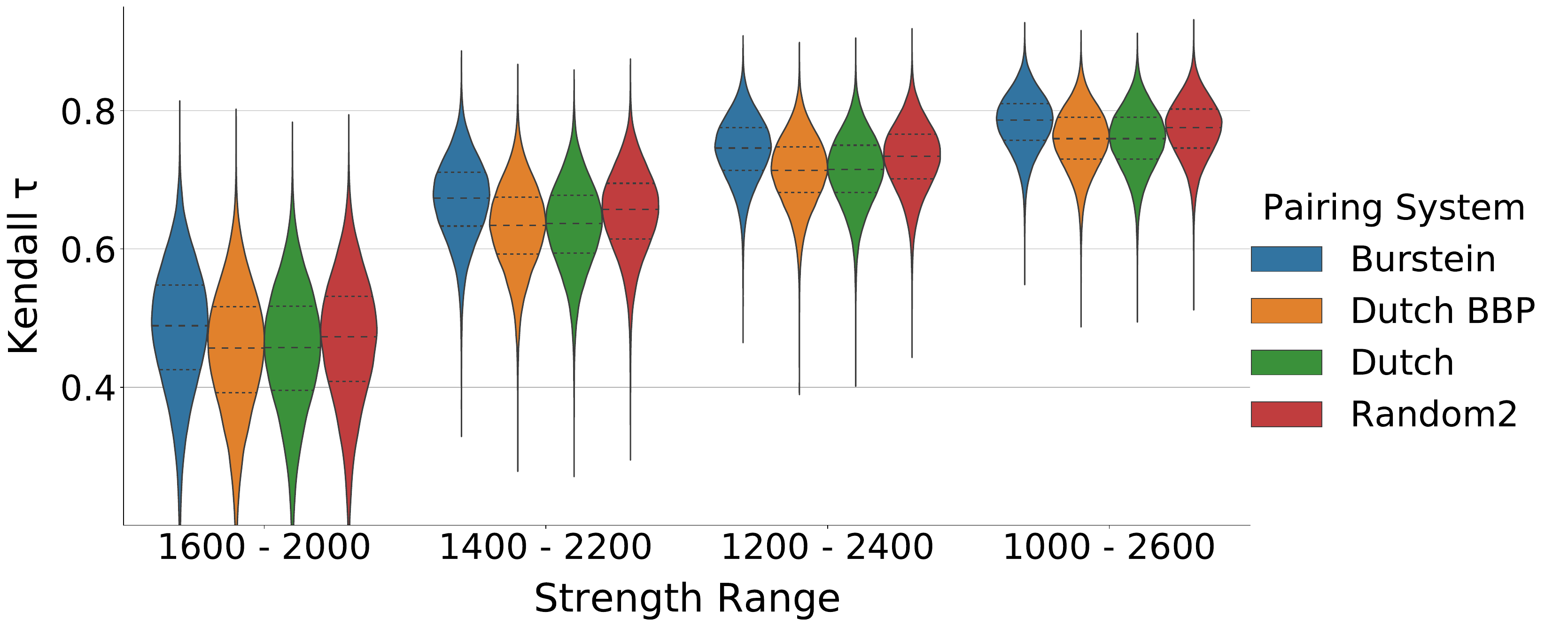}
    \caption{Ranking quality measured by normalized Kendall $\tau$ for different strength range sizes.}
    \label{fig:strength_range_size}
\end{figure} 
The results depicted in Figure~\ref{fig:strength_range_size}
show that also for different strength range sizes the MWM engine with Burstein or Random2 outperforms Dutch BBP in terms of ranking quality and that Dutch is on a par with Dutch BBP.

A higher strength range size results in higher ranking quality and less variance.
The increasing ranking quality can be explained by a higher mean strength difference, 
which results from a larger strength range size. Variance decreases, because match results become more predictable.

The difference in ranking quality between Burstein and Dutch BBP is much higher for a strength range size of 400 compared to 800 and 1200. For small strength range sizes in all Dutch BBP paired matches it is more likely that a weaker player wins against a stronger opponent, while for Burstein at least some matches are still predictable.

\subsection{Different Player Strength Distributions}

We provide additional experimental results that indicate that our findings hold independently of the employed player strength distributions, i.e., we get the same behavior also for non-uniform distributions. Since no data is available that let's us estimate how realistic player strength distributions look like, we focus on several natural candidates that deviate strongly from uniform distributions. 

\begin{figure}[ht]
    \centering
    \includegraphics[width=0.6\linewidth]{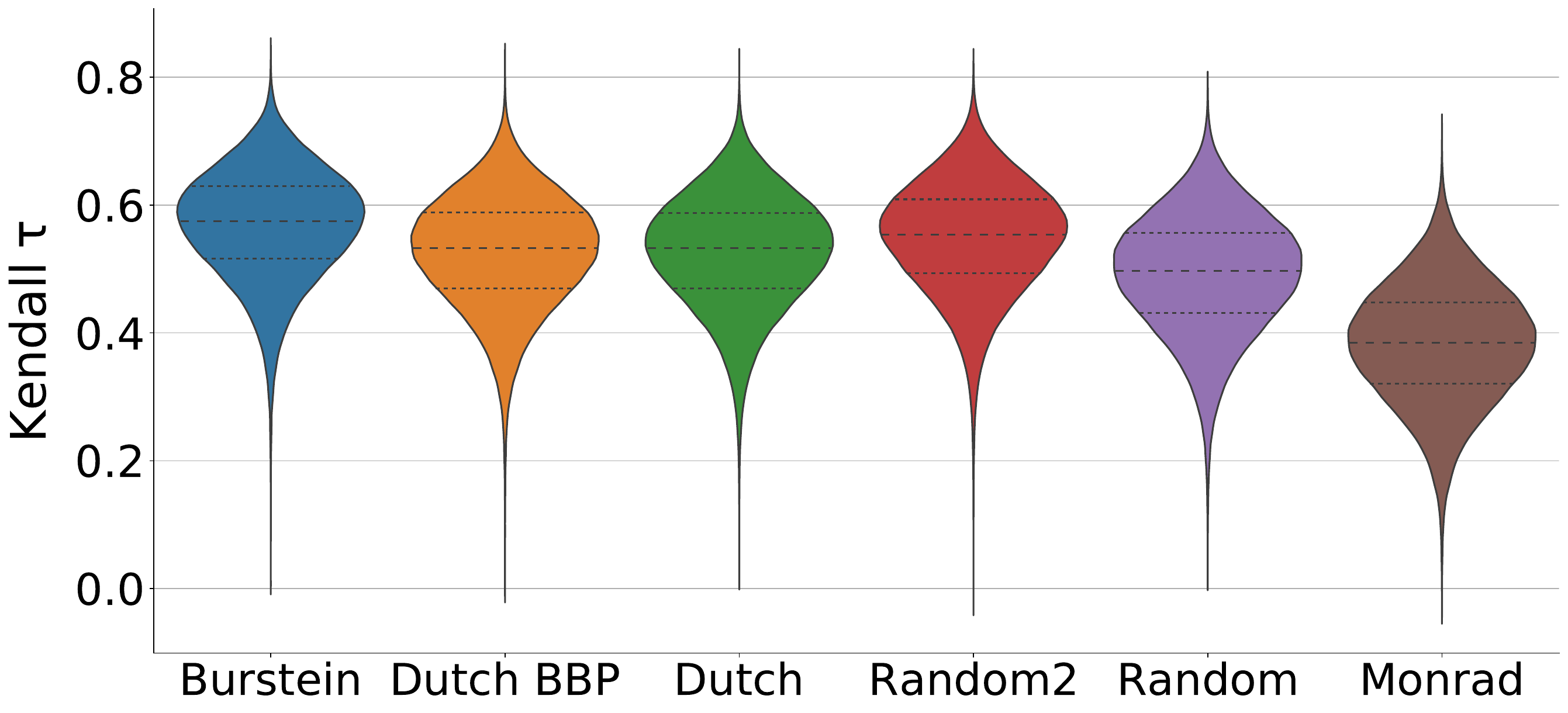}
    \caption{Ranking quality measured by normalized Kendall $\tau$ for 32 players with an exponential player strength distribution in the range $[1400,2200]$ with mean at $2000$. 
    }
    \label{fig:ranking_quality_exp_strong}
\end{figure}

\begin{figure}[ht]
    \centering
    \includegraphics[width=0.6\linewidth]{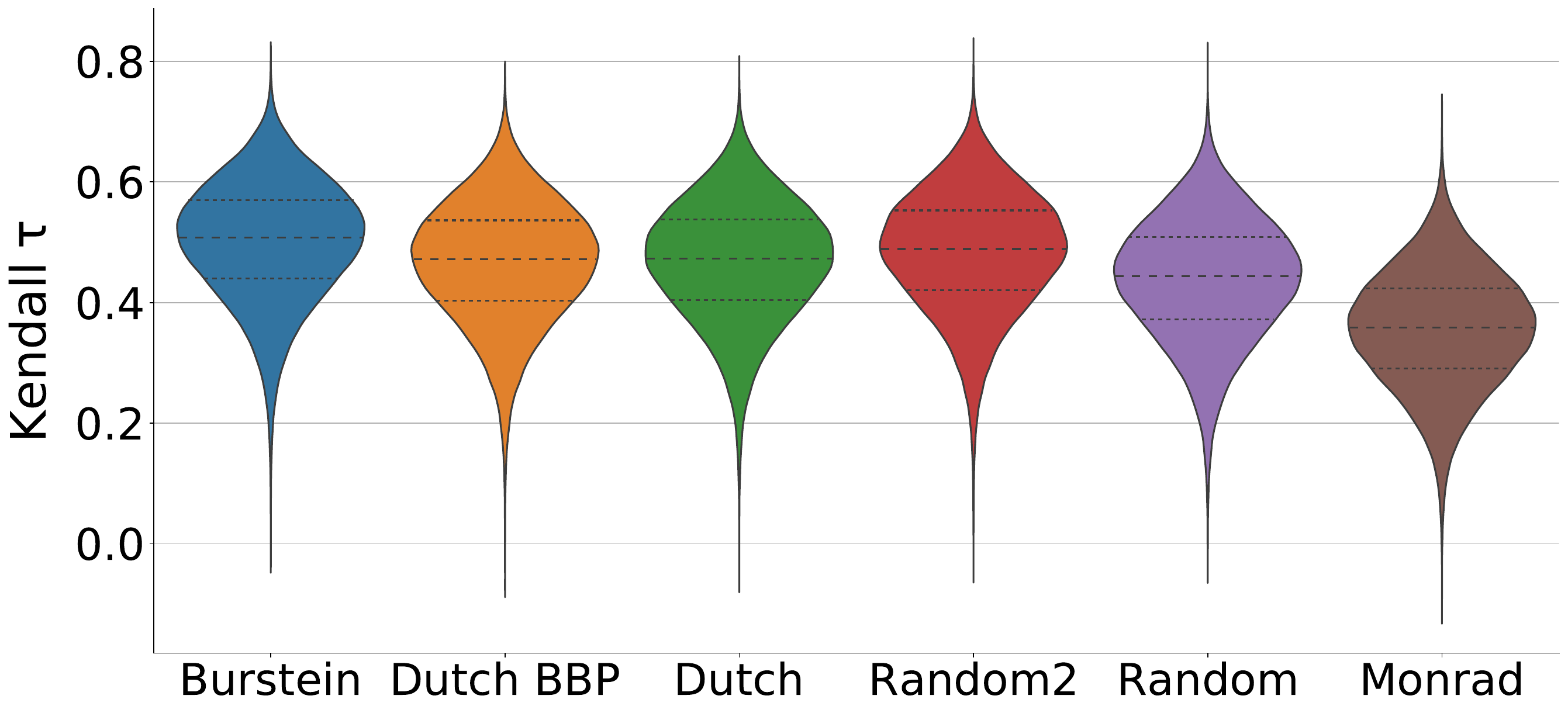}
    \caption{Ranking quality measured by normalized Kendall $\tau$ for 32 players with an exponential player strength distribution in the range $[1400,2200]$ with mean at $1600$. 
    }
    \label{fig:ranking_quality_exp_weak}
\end{figure}

\begin{figure}[h!]
    \centering
    \includegraphics[width=0.6\linewidth]{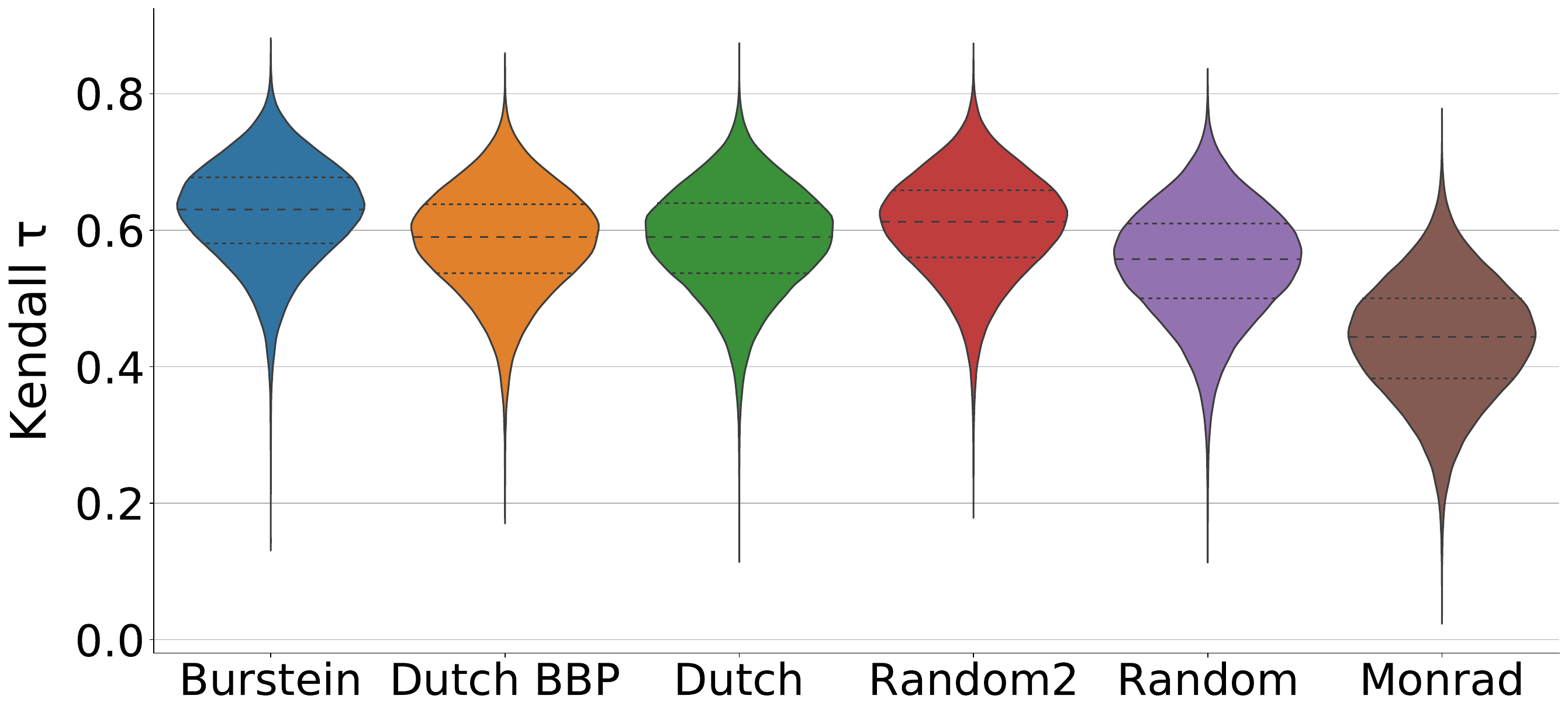}
    \caption{Ranking quality measured by normalized Kendall $\tau$ for 32 players with a normally distributed player strength distribution in the range $[1400,2200]$ with mean at $1800$ and standard deviation of $200$. 
    }
    \label{fig:ranking_quality_normal}
\end{figure}

First, we considered player strength distributions that are derived from exponential distributions. For this, we consider in Figure~\ref{fig:ranking_quality_exp_strong} a case with many strong players and only a few weak players and in Figure~\ref{fig:ranking_quality_exp_weak} a case with many weak players and only a few strong players within the given strength range size. 
We also considered player strength distributions derived from a normal distribution with a mean exactly in the middle of the strength range size and a standard deviation of a fourth of the strength range size. See Figure~\ref{fig:ranking_quality_normal} for the corresponding results.

\begin{figure}[h!]
    \centering
    \includegraphics[width=0.6\linewidth]{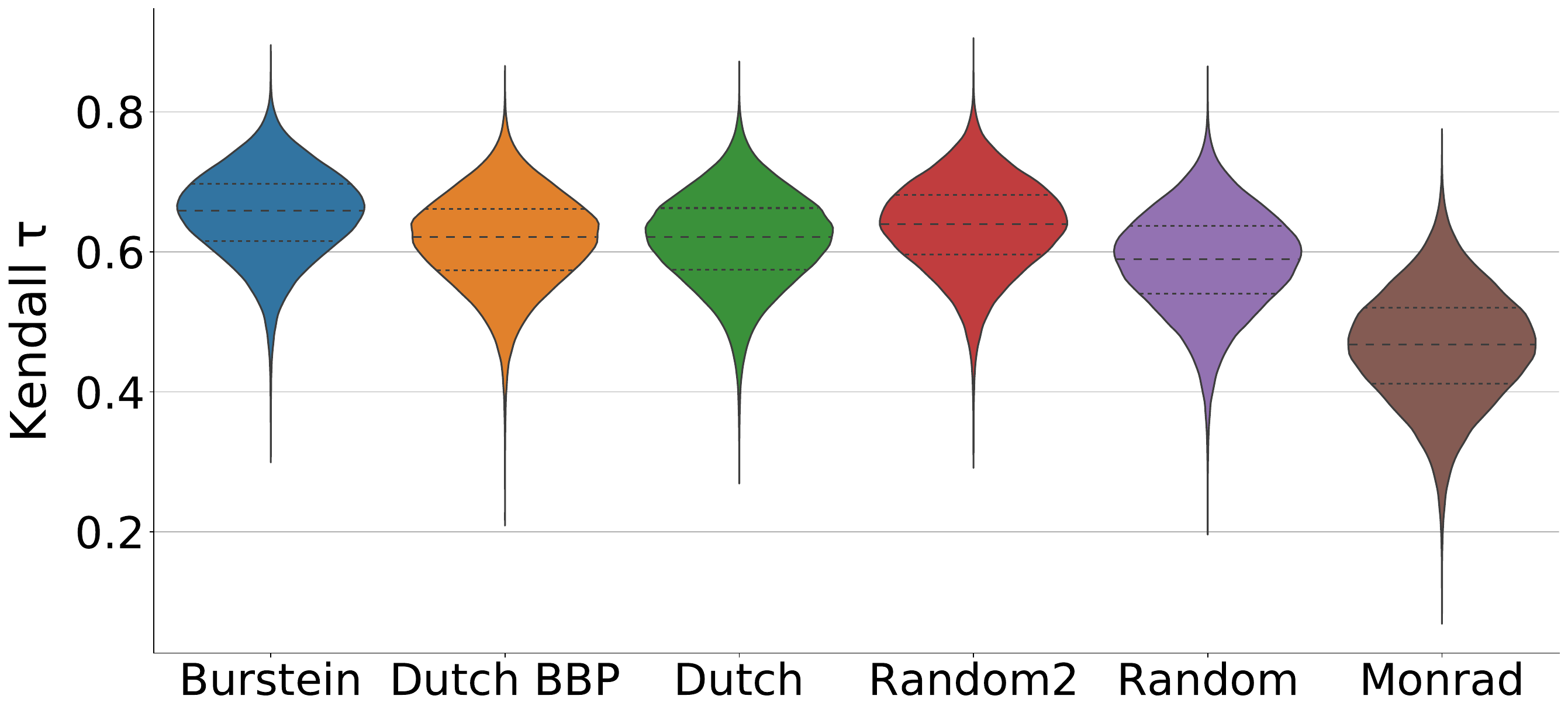}
    \caption{Ranking quality measured by normalized Kendall $\tau$ for 32 players uniformly sampled from the real-world distribution of Elo scores restricted to the range $[1400,2200]$. 
    }
    \label{fig:ranking_quality_realworld}
\end{figure}

Finally, we investigated a player strength distribution that is derived from uniformly sampling player strengths from the real-world distribution of Elo scores of all 363\,275 players listed by FIDE\footnote{See \url{https://ratings.fide.com/download_lists.phtml} for details.}, restricted to the desired strength range. Figure~\ref{fig:ranking_quality_realworld} shows also very similar results for this case.

\subsection{Ranking Quality via Spearman $\rho$ and NDCG}
For comparison reasons, we provide an evaluation of the achieved ranking quality via the Spearman $\rho$ and the normalized discounted cumulative gain (NDCG) measures. 

Besides Kendall $\tau$, Spearman $\rho$ is commonly used for comparing rankings. Here, we use a normalized variant of Spearman $\rho$, similar to the normalized Kendall $\tau$. 

The NDCG measure is not commonly used for comparing rankings. It is used to evaluate search engines, by assigning a relevance rating to documents and awarding a higher score if highly relevant documents are listed early. Applied to our case, NDCG puts an emphasis on ranking the top players correctly, while ranking the lowest ranked players correctly is basically irrelevant.

\begin{figure}[h!]
    \centering
    \includegraphics[width=0.6\linewidth]{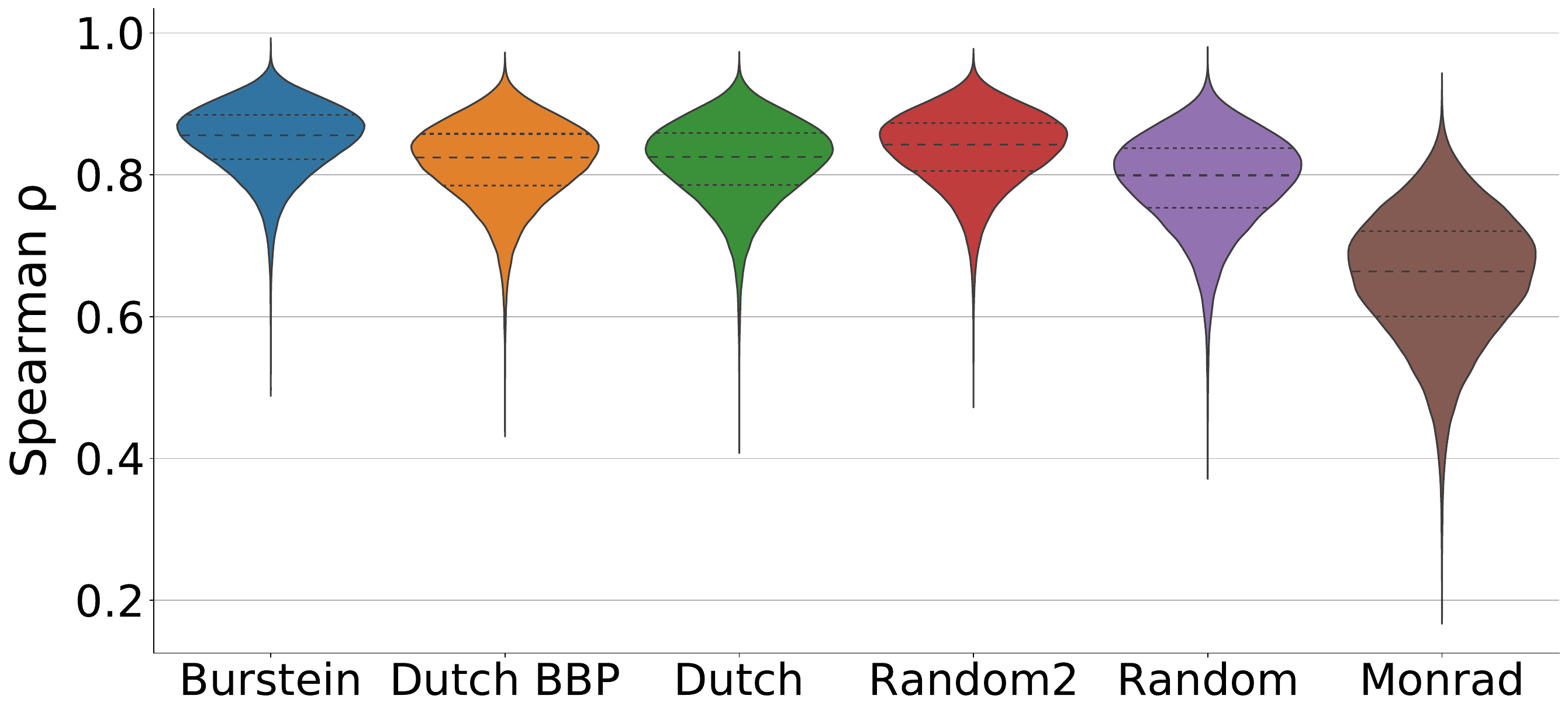}
    \caption{Ranking quality measured by normalized Spearman $\rho$. 
    }
    \label{fig:rho_ranking_quality}
\end{figure}

\begin{figure}[h!]
    \centering
    \includegraphics[width=0.6\linewidth]{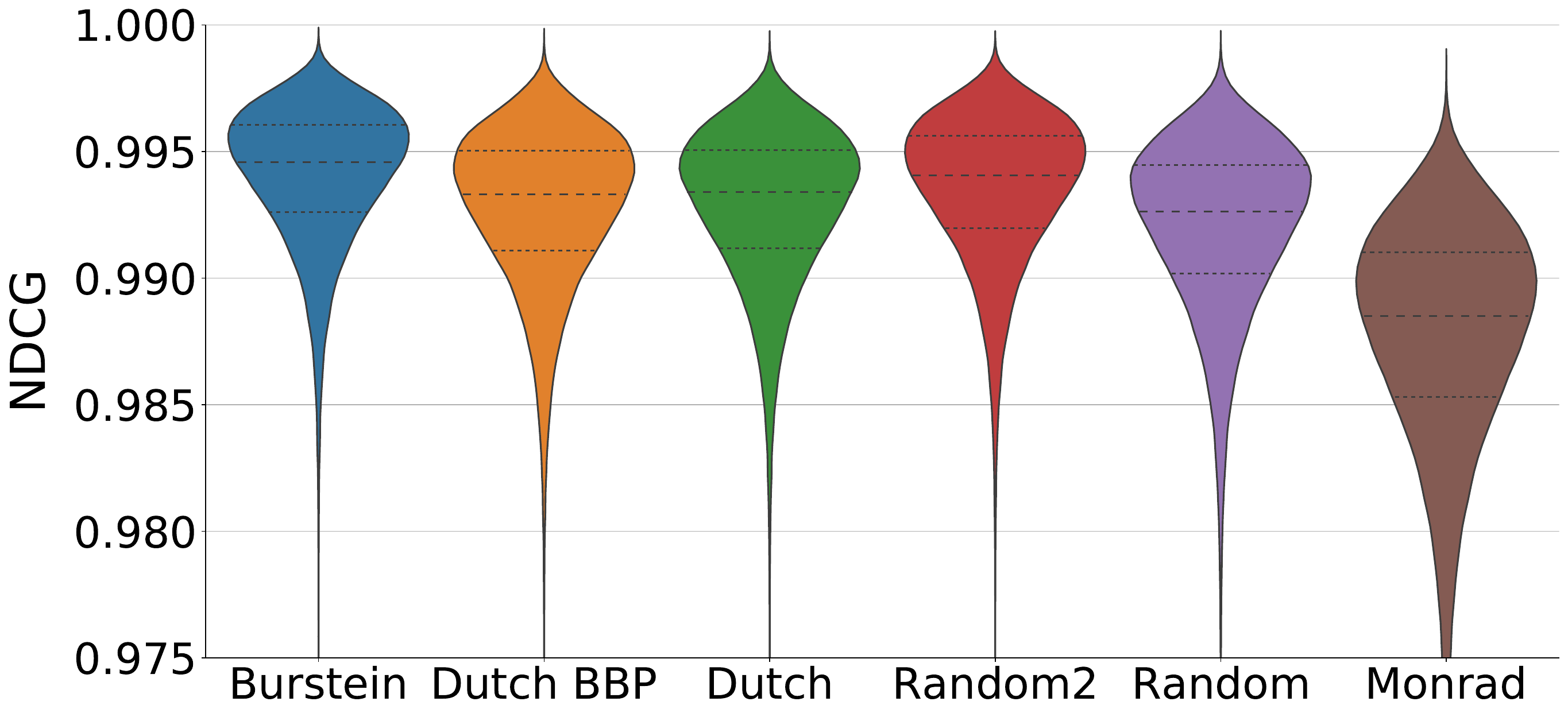}
    \caption{Ranking quality measured by the normalized discounted cumulative gain (NDCG). 
    }
    \label{fig:ndcg_ranking_quality}
\end{figure}

\begin{figure}[h!]
    \centering
    \includegraphics[width=0.7\linewidth]{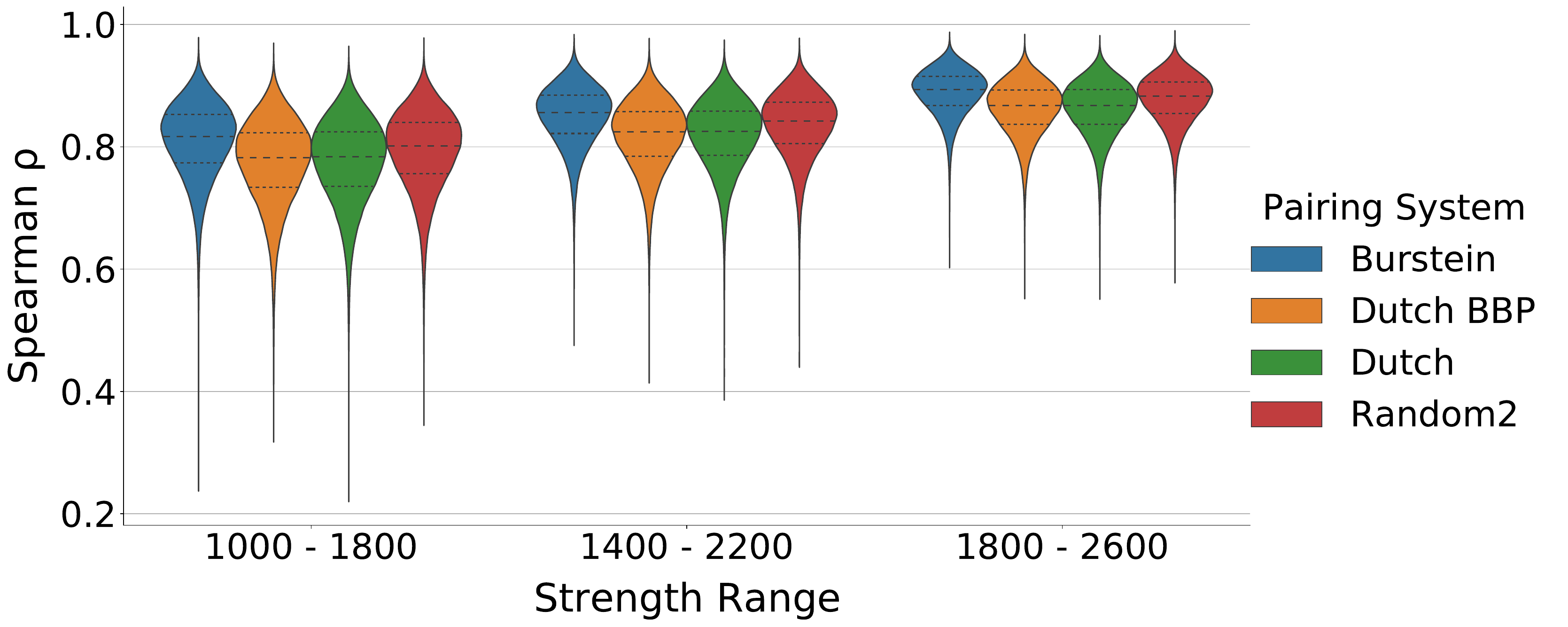}
    \caption{Ranking quality measured by normalized Spearman $\rho$.}
    \label{fig:rho_mean_strength}
\end{figure}

\begin{figure}[h!]
    \centering
    \includegraphics[width=0.7\linewidth]{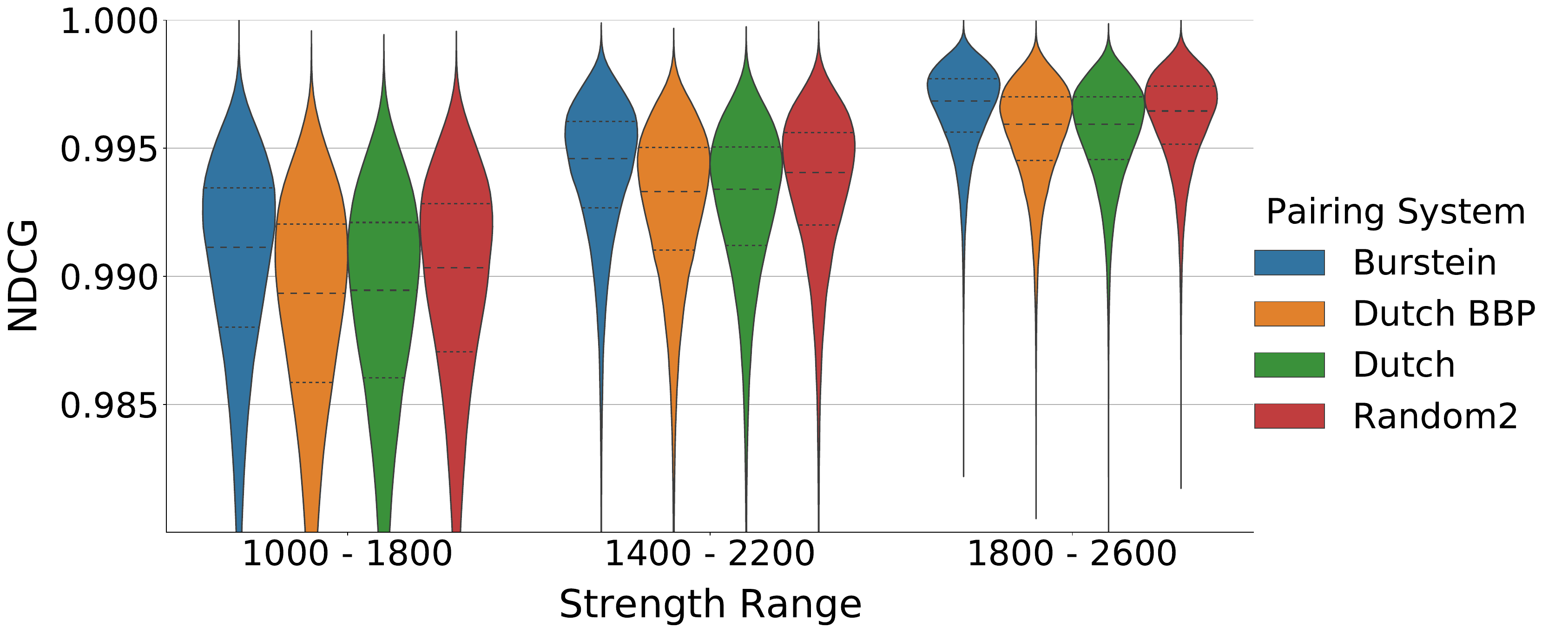}
    \caption{Ranking quality measured by the normalized discounted cumulative gain (NDCG).}
    \label{fig:ndcg_mean_strength}
\end{figure}

\begin{figure}[h!]
    \centering
    \includegraphics[width=0.7\linewidth]{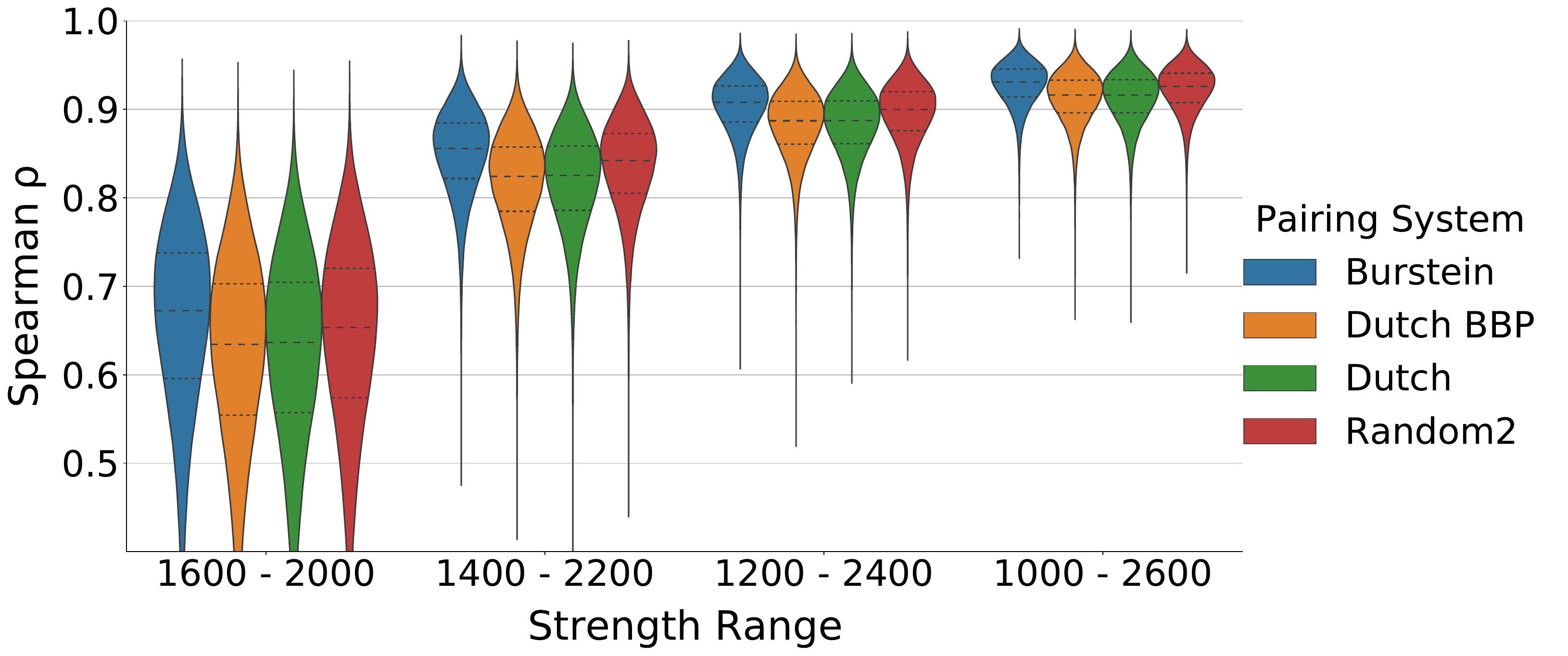}
    \caption{Ranking quality measured by normalized Spearman $\rho$.}
    \label{fig:rho_strength_range_size}
\end{figure}

\begin{figure}[h!]
    \centering
    \includegraphics[width=0.7\linewidth]{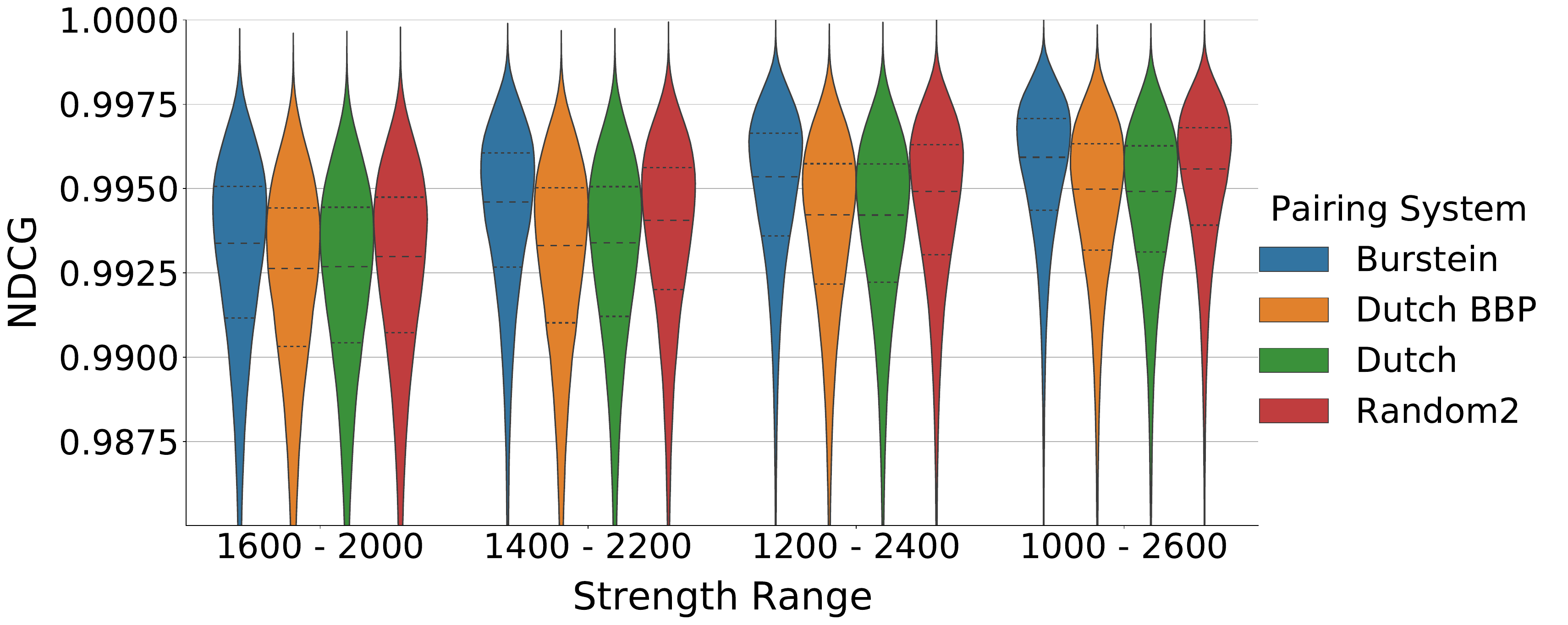}
    \caption{Ranking quality measured by the normalized discounted cumulative gain (NDCG).}
    \label{fig:ndcg_strength_range_size}
\end{figure}

As shown in Figure~\ref{fig:rho_ranking_quality} and Figure~\ref{fig:ndcg_ranking_quality}, the results with normalized Spearman $\rho$ and NDCG look almost identical to the results for normalized Kendall $\tau$ in Figure~\ref{fig:ranking_quality}.
\noindent Also, for different strength ranges or range sizes we get consistent results, see Figures~\ref{fig:rho_mean_strength}, \ref{fig:ndcg_mean_strength}, \ref{fig:rho_strength_range_size} and \ref{fig:ndcg_strength_range_size}.

\pagebreak 

\section{Fairness}
Here we present additional simulation results that measure the achieved fairness, i.e., results regarding the compliance with the quality criteria (Q1) and (Q2).
\subsection{Number of Float Pairs}
\begin{figure}[ht]
    \centering
    \includegraphics[width=0.7\linewidth]{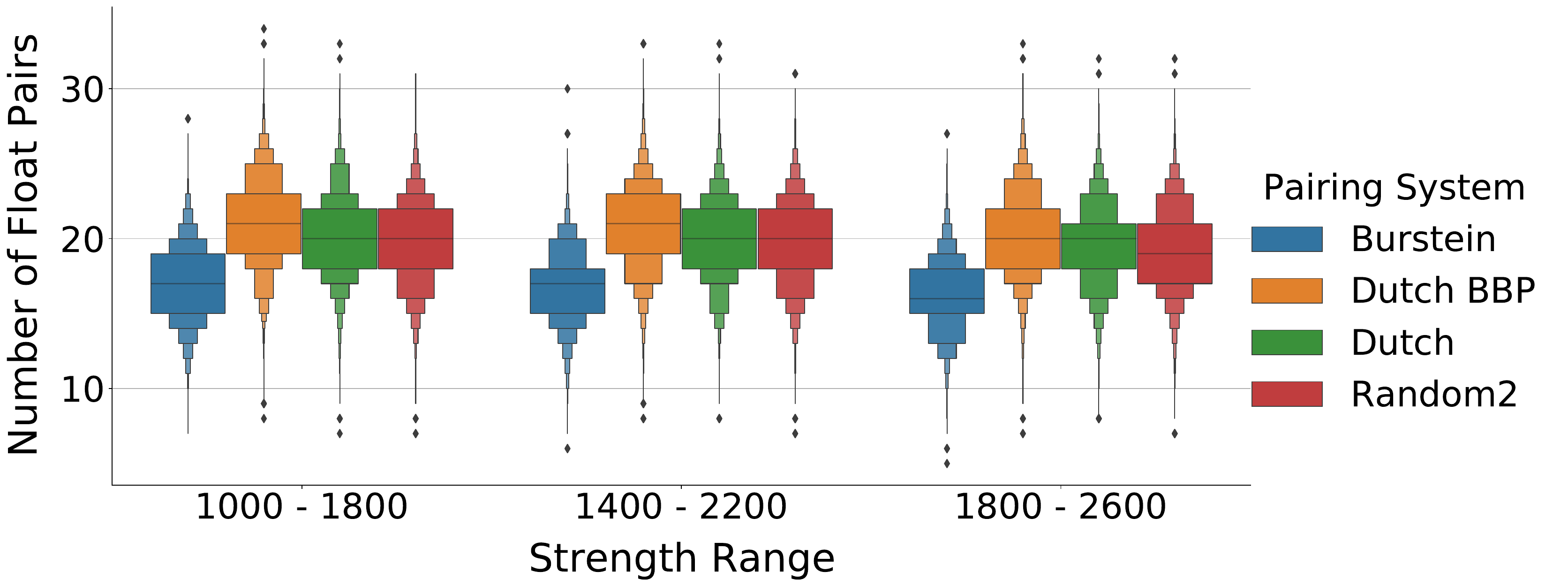}
    \caption{Number of float pairs for different strength ranges.}
    \label{fig:num_float_pairs_strength_ranges}
\end{figure}

\begin{figure}[ht]
    \centering
    \includegraphics[width=0.7\linewidth]{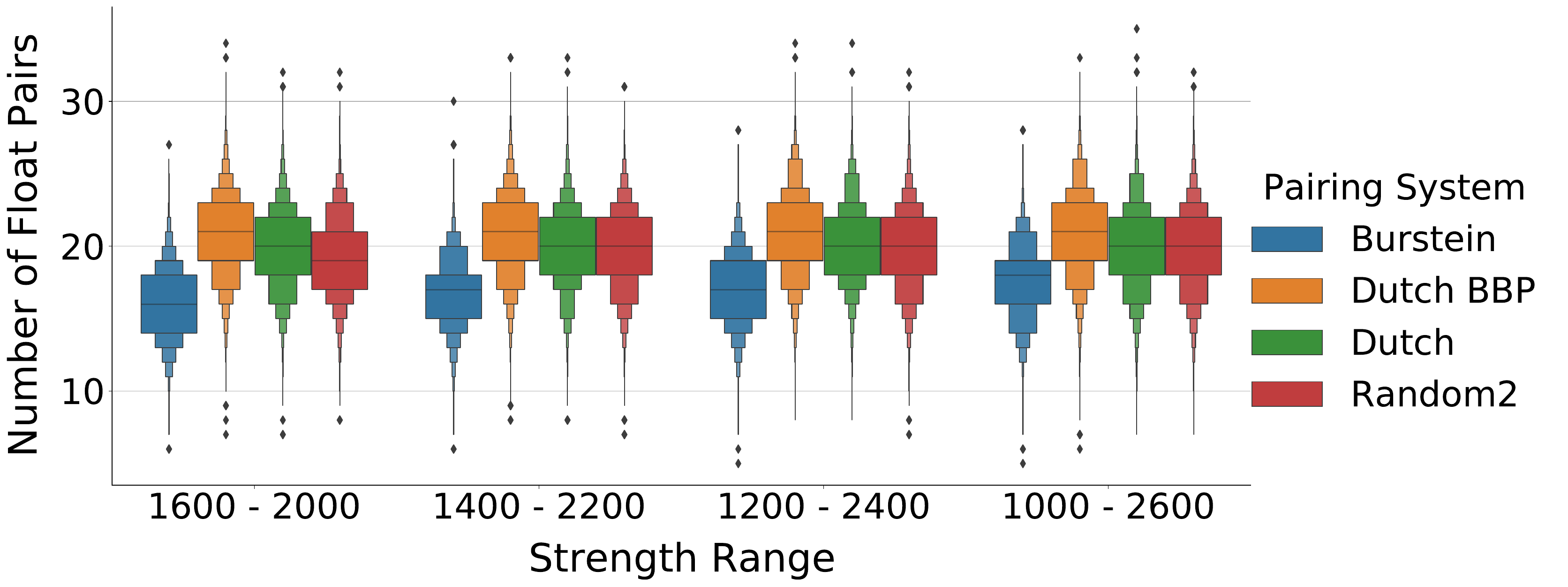}
    \caption{Number of float pairs for different strength range sizes.}
    \label{fig:num_float_pairs_strength_range_sizes}
\end{figure}

We consider the obtained number of float pairs for different strength ranges and different strength range sizes. Figures~\ref{fig:num_float_pairs_strength_ranges} and \ref{fig:num_float_pairs_strength_range_sizes} show that we get consistent results for different strength ranges and different strength range sizes. Burstein has by far the lowest number of float pairs, but also Random2 and Dutch perform slightly better than Dutch BBP. 
 
\begin{figure}[h!]
    \centering
    \includegraphics[trim={0 0 115mm 0}, clip,width=0.6\linewidth]{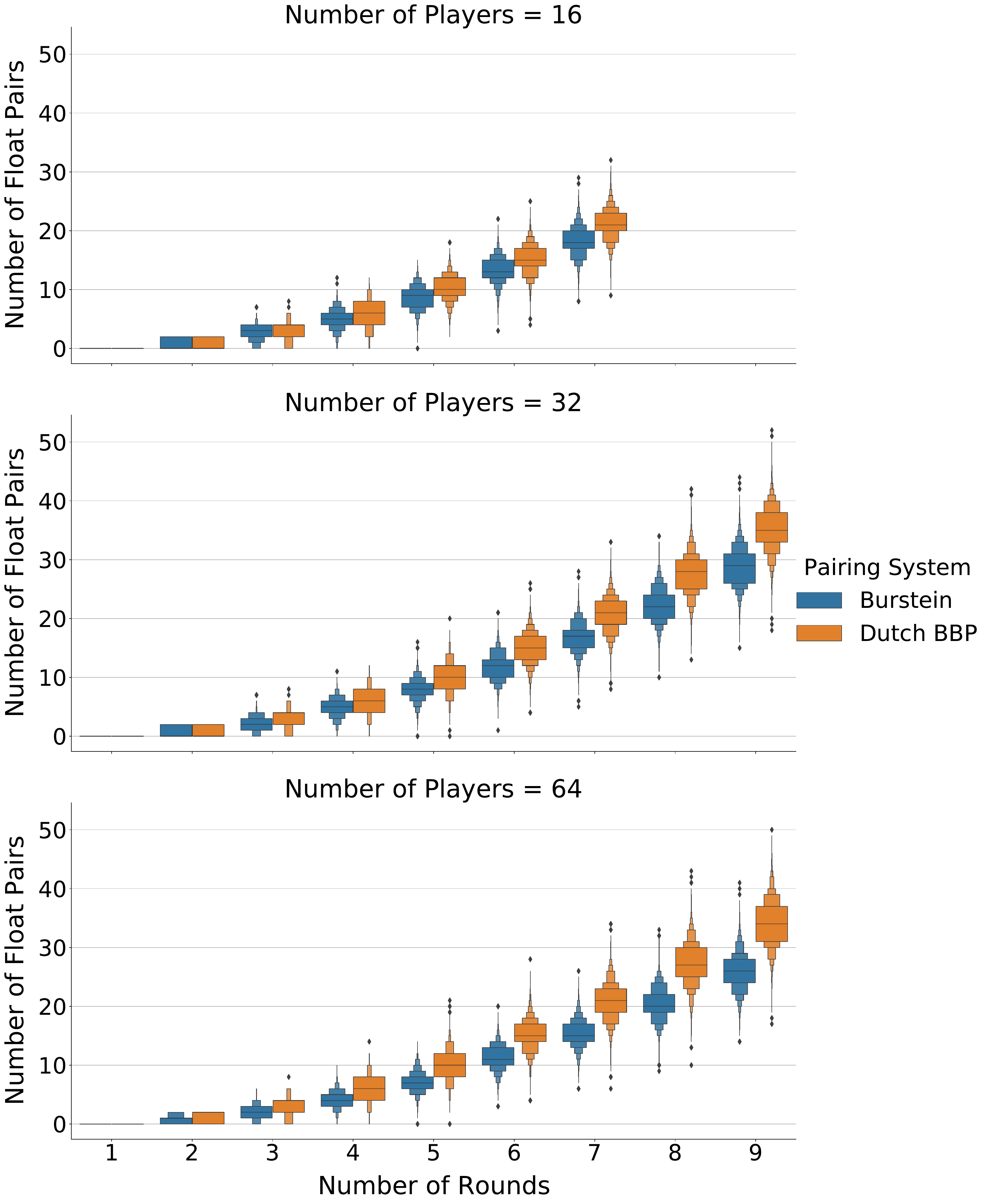}
    \caption{Number of float pairs for different tournament sizes and lengths. The results for Burstein are shown in blue, results for Dutch BBP in orange.}
    \label{fig:num_float_pairs_num_players_rounds}
\end{figure}

\noindent Figure~\ref{fig:num_float_pairs_num_players_rounds} shows a direct comparison of the obtained number of float pairs for Burstein and Dutch BBP for different numbers of players and different tournament lengths. 

Also here we consistently get that Burstein achieves much fewer float pairs than Dutch BBP.



\subsection{Absolute Color Difference}
\begin{figure}[ht]
    \centering
    \includegraphics[trim={0 0 114mm 0},clip,width=0.6\linewidth]{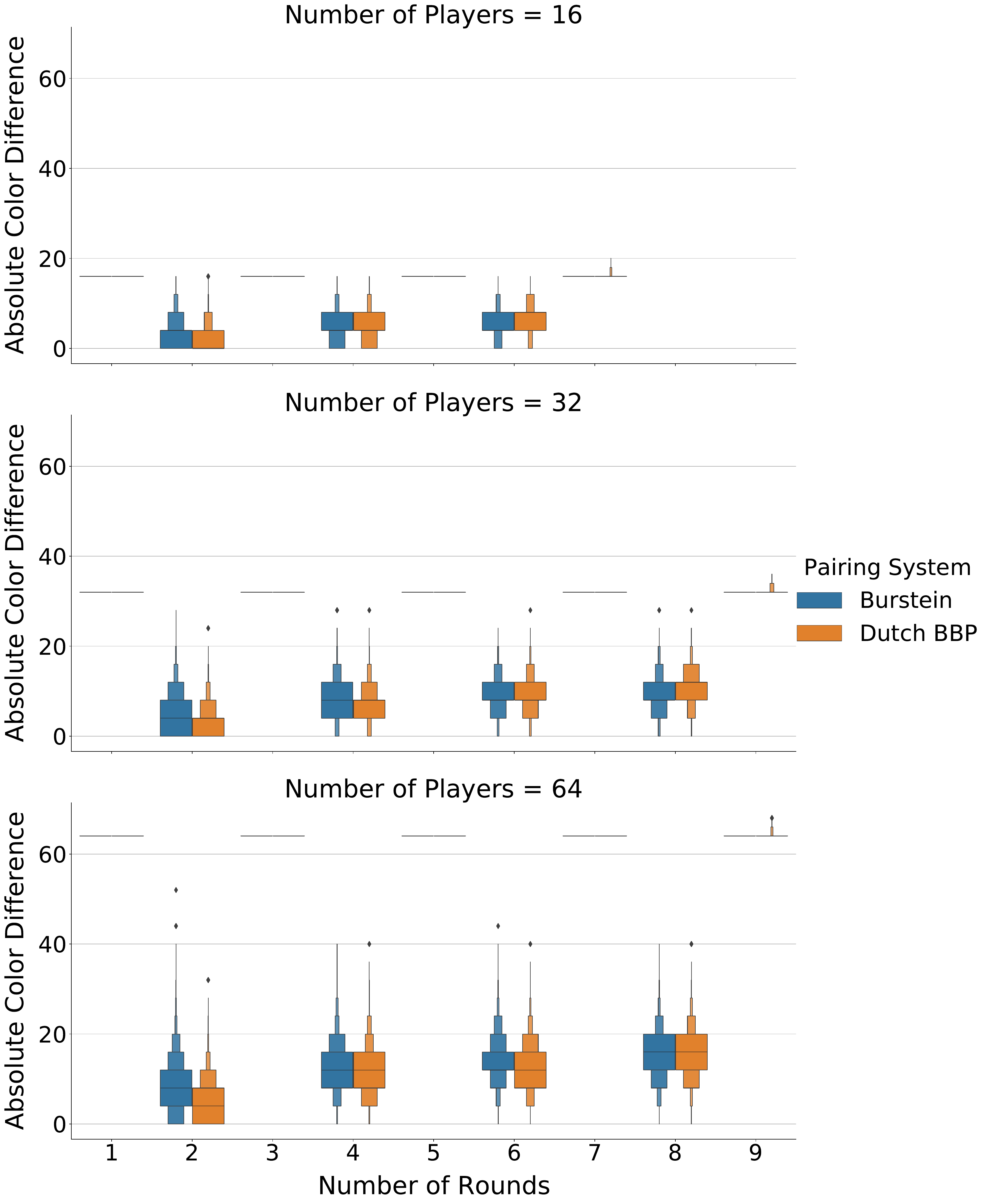}
    \caption{Absolute color difference in rounds 1-9, 16-64 players with strength range 1400-2200. Results for Burstein are shown in blue, Dutch BBP results are shown in orange.}
    \label{fig:color_diff_number_of_players_rounds}
\end{figure}

The measured absolute color difference increases slightly with the number of rounds and also with the number of players, as Figure~\ref{fig:color_diff_number_of_players_rounds} shows.

Note that in every odd round, the absolute color difference must be at least $n$, which can also be seen. All investigated pairing systems almost always meet this lower bound for odd rounds. Interestingly, Dutch BBP seems to perform slightly better in tournaments with at most 4 rounds compared to Burstein, but this tiny advantage vanishes for at least six rounds. We get similar results when comparing with Random2, Dutch, Random, and Monrad. 

\end{document}